\documentclass[aps,pre,amsmath,amsfonts,amssymb,superscriptaddress,preprint]{revtex4}
\usepackage[latin1]{inputenc}
\usepackage[pdftex]{graphicx}

\usepackage{relsize}
\usepackage{tikz}
\usepackage{amsmath,latexsym,mathrsfs, amssymb}
\usetikzlibrary{decorations.markings}

\everymath{\displaystyle}
\newcommand{\beq}{\begin{equation}}
\newcommand{\eeq}{\end{equation}}

\newcommand{\bi}{\begin{itemize}}
\newcommand{\ei}{\end{itemize}}
\newcommand{\D}{\mathrm{d}}
\newcommand{\bra}[1]{\left\langle #1 \right|}
\newcommand{\ket}[1]{\left| #1 \right\rangle}
\newcommand{\Adj}[1]{\widehat{#1}^{\dag}}
\newcommand{\braket}[2]{\left\langle #1  \Big\vert  #2  \right\rangle}

\newcommand{\affA}{Aix-Marseille University, Marseille, France}
\newcommand{\affB}{CNRS Centre de Physique Th\'eorique UMR7332,
13288 Marseille, France}
\newcommand{\affC}{Centre d'Immunologie de Marseille-Luminy,
13288 Marseille, France}

\newcommand{\affG}{Department of Physics and Department of Oncology, 3-336, Cross Cancer Institute, 
Edmonton, AB, T6G 1Z2, Canada}
\newcommand{\affF}{IES, University of Montpellier -- CNRS, UMR 5214, 34000 Montpellier, France}
\newcommand{\affH}{Department of Physics, University of Rome  "La Sapienza", Italy}
\newcommand{\affL}{Laboratoire d'Ing\'enierie des Syst\`emes Macromol\'eculaires UMR7255, 13402 Marseille, France}

\begin{document}
\DeclareGraphicsExtensions{.jpg,.pdf,.mps,.png,.ai}

\title{Out-of-equilibrium collective oscillation as phonon condensation in a model protein}

\author{Ilaria Nardecchia}
\email{i.nardecchia@gmail.com}
\affiliation{\affA}\affiliation{\affB}\affiliation{\affC}
\author{Jeremie Torres}
\email{jeremie.torres@umontpellier.fr}
\affiliation{\affF}
\author{Mathias Lechelon}
\email{mathias.lechelon@gmail.com}
\affiliation{\affA}\affiliation{\affB}\affiliation{\affC}

\author{Valeria Giliberti}
\email{valeria.giliberti@roma1.infn.it}
\affiliation{\affH}
\author{Michele Ortolani}
\email{michele.ortolani@roma1.infn.it}
\affiliation{\affH}
\author{Philippe Nouvel}
\email{philippe.nouvel@umontpellier.fr}
\affiliation{\affF}
\author{Matteo Gori}
\email{gori6matteo@gmail.com}
\author{Yoann Meriguet}
\email{yoann.meriguet@umontpellier.fr}
\affiliation{\affF}
\author{Irene Donato}
\email{ irene.irened@gmail.com}
\affiliation{\affA}
\affiliation{\affB}
\author{Jordane Preto}
\email{jordane.preto@gmail.com}
\affiliation{\affG}
\author{Luca Varani}
\email{luca.varani@umontpellier.fr}
\affiliation{\affF}
\author{James Sturgis}
\email{sturgis@imm.cnrs.fr}
\affiliation{\affA}\affiliation{\affL}
\author{Marco Pettini}
\email{pettini@cpt.univ-mrs.fr; corresponding_author}
\affiliation{\affA}\affiliation{\affB}
\begin{abstract}
In the first part of the present paper (theoretical), the activation of out-of-equilibrium collective oscillations of a macromolecule is described as a classical phonon condensation phenomenon. If a macromolecule is modeled as an open system, that is, it is subjected to an external energy supply and is in contact with a thermal bath to dissipate the excess energy, the internal nonlinear couplings among the normal modes make the system undergo a non-equilibrium phase transition when the energy input rate exceeds a threshold value. This transition takes place between a state where the energy is incoherently distributed among the normal modes, to a state  where the input energy is channeled into the lowest frequency mode entailing a coherent oscillation of the entire molecule. The model put forward in the present work is derived as the classical counterpart of a quantum model proposed long time ago by H. Fr\"ohlich in the attempt to explain the huge speed of enzymatic reactions.
In the second part of the present paper (experimental),  we show that such a phenomenon is actually possible.
Two different and complementary THz near-field spectroscopic techniques, a plasmonic rectenna, and a micro-wire near-field probe, have been used in two different labs to get rid of artefacts.
By considering a aqueous solution of a model protein, the BSA (Bovine Serum Albumin),  we found that  this protein displays a remarkable absorption feature around 0.314 THz, when driven in a stationary out-of-thermal equilibrium state by means of optical pumping. 
The experimental outcomes are in very good qualitative agreement with the theory developed in the first part, and in excellent quantitative agreement with a theoretical result allowing to identify the observed spectral feature with a collective oscillation of the entire molecule.
\end{abstract}
\vfil
\date{\today}
\maketitle

\section{Introduction}
\textcolor{black}{Recent progress in terahertz technology has enabled to look at biological systems with terahertz radiation, that is, in an energy domain (a few meV) which is of the order of the activation energy of many biological processes. These include the excitation of collective modes of vibration of biomolecules for which experimental evidence has been originally provided  {\it at thermal equilibrium} by means of Raman  spectroscopy \cite{painter} and, recently, by several terahertz spectroscopic studies mainly carried on using dry or low-hydrated powders because of the very strong absorption of water \cite{turton,acbas,falconer,markelz}. More recent studies also addressed solvated proteins \cite{xu,ebbinghaus}. Collective oscillations of biomolecules, possibly driven by metabolic activity in living matter, by bringing about large oscillating dipole moments,  could activate resonant (thus selective), intermolecular attractive  electrodynamic forces acting at a long distance. But we have shown \cite{Preto:2015} that this is not possible at thermal equilibrium, where all the mentioned experiments have been hitherto performed. Thus, understanding whether long-range intermolecular electrodynamic forces can be really activated requires to figure out if out-of-thermal equilibrium collective oscillations of biomolecules can be realised. The possibility of activating these forces could help explaining  the fast encounters of the cognate partners of  biochemical reactions in living matter, even in the context of low molecular concentration of one of the reactants, as it might be the case of some ligand-receptor reaction or of the transcription machinery of the DNA.  From a theoretical viewpoint, collective oscillations of macromolecules have been described by H. Fr\"ohlich \cite{Frohlich:1968,Froeh:1970,Froeh:1972,Frohlich:1977}  as a Bose-like condensation of their normal vibrational modes. }

However,  Fr\"ohlich's theory has been criticized and marginalized  because of several reasons. Among the others  the original  theory is certainly oversimplified and lacks many important explanations to qualify as really relevant to understand basic mechanisms of life. 
Recently, a remarkable progress was done in the direction of assessing the possible relevance of Fr\"ohlich condensates in a revisitation of the theory \cite{reimers} which has led to identify three different kinds of Fr\"ohlich condensates: weak, strong, and coherent. The authors explain that weak condensates may have profound effects on chemical and enzyme kinetics, while coherent condensates are shown to involve very large amounts of energy, to be very fragile and not to be produced by the Wu-Austin Hamiltonian from which Fr\"ohlich's rate equations can be derived. The work of Ref.\cite{reimers} is still theoretical and developed in a quantum framework.

From the experimental side, very interesting observations of phonon condensates of quantum nature at thermal equilibrium have been recently reported in the literature.  Remarkably, optical phonon condensates have been observed in heterostructures at room temperature \cite{altfeder}, arising from phase and frequency dynamical synchronisation of phonons due to suitable scattering processes around defects. Another intriguing result  concerns polariton Bose condensation at thermal equilibrium in a GaAs high-Q micro cavity in a temperature range of $10 - 25 K$ \cite{polaritons}, still a quite high temperature with respect to Bose-Einstein condensation of alkali atoms. These nice results concern optical phonons or polaritons weakly interacting with phonons, respectively, in condensed matter systems. But what about low frequency vibrations of  biomolecules at room temperature? We can wonder whether a further and complementary step forward can be done going beyond the quantum treatment of Fr\"ohlich condensates, also after the revisitation of Ref.\cite{reimers}. 

In fact, the frequency of collective oscillations of bio-macromolecules is expected in the sub-THz domain, around $10^{11}$ Hz, hence at room temperature it is $k_BT/\hbar\omega\gg 1$ (where $k_B$ and $\hbar$ are the Boltzmann and Planck's constants respectively)  and the average number of phonons estimated at this frequency with the Bose-Einstein statistics  ($\langle n\rangle\simeq 62.01$) is the same as that given by the Boltzmann statistics ($\langle n\rangle\simeq 62.51$) within a $1\%$ accuracy, as a consequence a classical description seems reasonable. Moreover, when the system is brought out of thermal equilibrium by external energy injection the effective temperature of each normal mode is higher, thus increasing the occupation number of each mode and making the classical approximation even more reasonable. Furthermore, for example, the vibrational properties of proteins are very well described by Molecular Dynamics simulations performed in a classical context. 
\textcolor{black}{Therefore, the first important question addressed in the present work is whether a phonon condensation phenomenology can be retrieved also in a classical framework and out of thermal equilibrium.  
To this aim, by resorting to a de-quantization method, we have worked out a classical version of the original Fr\"ohlich model, finding that - remarkably -  in a classical context too a Bose-like phonon condensation is possible \cite{classicalcondens}. This possibility requires considering a biomolecule as an open system - that is, far from thermal equilibrium with its environment} -  through which energy flows under the simultaneous actions of an external energy supply and of dissipation due to radiative, dielectric, and viscous energy losses.  \textcolor{black}{We have found that the classical Bose-like condensation in the lowest vibrational mode occurs when the energy input rate exceeds some threshold value. Then, this a-priori non obvious result has motivated an experimental effort to find out if the theoretically predicted phenomenon is actually possible in the physical world. Two independent experiments, in geographically distinct laboratories, have confirmed the existence of out-of-equilibrium collective oscillations for a model protein. This is a proof of concept of which the most significant implication is that, in compliance with a theoretical prediction \cite{Preto:2015},  a crucial pre-requisite is fulfilled to excite intermolecular long-range electrodynamic interactions. In their turn, as already said above, these interactions could affect the biomolecular dynamics by contributing to drive the high efficiency and rapidity of mutual encounters of the partners of biochemical reactions in living matter, encounters which sometimes could hardly appear to be the result of Brownian diffusion only. }

The derivation of the mentioned classical model - together with the numerical evidence of phonon condensation - is reported in Section \ref{classicBose}. The results of two complementary experiments are reported and discussed in Section \ref{manips}. Materials and methods are described in Section \ref{materials}. In Section \ref{conclusions} we draw some conclusion.

\section{Classical out-of-equilibrium phonon condensation}\label{classicBose}
Some decades ago, the study of open systems far from thermodynamic equilibrium showed, under suitable conditions, the emergence of self-organization. Striking similarities were observed among very different physical systems having in common the fact of being composed of many non-linearly interacting subsystems. When a control parameter, typically the energy input rate, exceeds a critical value (that is, when the energy gain exceeds the energy losses) then the subsystems act cooperatively to self-organize in what is commonly referred to as a non-equilibrium phase transition \cite{haken1,haken2}. This phenomenon is a generic consequence of the presence of the mentioned basic ingredients, that is,  nonlinearly interacting subsystems, dissipation to a thermal bath, and  external energy supply.  
Therefore,  even idealised models of real systems are capable of correctly catching the occurrence of these collective behaviors, at least qualitatively.
This fascinating topic was pioneered in the late '60s by H. Fr\"ohlich with the model mentioned in the Introduction. 
However, Fr\"ohlich rate equations for the vibrational mode amplitudes, as occupation numbers in Fock's representation, of a generic biomolecule were originally put forward heuristically: the original formulation was lacking a microscopic model.
 A microscopic Hamiltonian - from which the Fr\"ohlich's rate equations can be derived - was later given by Wu and Austin \cite{wu1,wu2,wu3,wu4} to model the dynamics of the normal modes of a macromolecule, of the thermal bath surrounding it, and of the external energy pump. The Wu-Austin Hamiltonian is formulated in second quantization formalism. 
 
 In what follows we indicate with $\Adj{a}_{\omega_i},\widehat{a}_{\omega_i}$ the quantum creation/annihilation operators for the vibrational normal modes of the main system (i.e. a biomolecule) with frequency $\omega_i\in\mathcal{I}_{sys}$.\\
Such a system is put in contact with a thermal bath which represents the degrees of freedom of the
environment surrounding the protein and, possibly, other normal modes of the protein which can be
considered at thermal equilibrium with the surrounding environment. The thermal bath is characterized by a
temperature $T_{B}$ and it is represented by a collection of harmonic oscillators with characteristic frequencies $\Omega_j\in\mathcal{I}_{bth}$ whose annihilation/creation operators are  $\widehat{b}_{\Omega_j}$ and $\Adj{b}_{\Omega_j}$, respectively.\\
In order to put the system representing normal modes of a biomolecule out of thermal equilibrium, an external source of energy is necessary: such an external source is represented as another thermal bath at a temperature $T_S \gg T_B$. Also in this case the corresponding thermal bath is described by a collection  of harmonic oscillators with frequencies $\Omega_{k}^{'}\in\mathcal{I}_{src}$,  the quantum annihilation/creation operators of which are $\widehat{c}_{\Omega_{k}^{'}}$ and $\Adj{c}_{\Omega_{k}^{'}}$. These three sets of harmonic oscillators can be regarded as three subsystems of a larger isolated system $\mathcal{S}$ (we coherently indicate with $\mathcal{I}_{\mathcal{S}}$ the set of all the normal modes of the system) whose quantum dynamics is described by the Hamiltonian
\begin{equation}
\label{eq:HtotQuant}
\widehat{H}_{Tot}=\widehat{H}_{0}+\widehat{H}_{Int}
\end{equation}
where $\widehat{H}_{0}$ is the free Hamiltonian of the three sets of harmonic oscillators representing the molecular 
normal modes and the two heat baths
\begin{equation}
\label{eq:HfreeQuan}
\widehat{H}_{0}=\sum_{\omega_i\in\mathcal{I}_{sys}}\,\hslash \omega_i\,\, \Adj{a}_{\omega_i}\widehat{a}_{\omega_i}+\sum_{\Omega_{j}\in\mathcal{I}_{bath}}\,\hslash \Omega_{j}\,\,\Adj{b}_{\Omega_j}\widehat{b}_{\Omega_j}+\sum_{\Omega_k^{'}\in\mathcal{I}_{src}}\,\hslash\Omega_k^{'}\,\,\Adj{c}_{\Omega_k^{'}}\widehat{c}_{\Omega_{k}^{'}}\,.
\end{equation}
The interactions among normal modes are described by $\widehat{H}_{Int}$; in the original formulation by Wu and Austin the interaction term has the form:
\begin{equation}
\label{eq:HintQuan_WuAustin}
\begin{split}
\widehat{H}_{IntWA}=&\widehat{H}_{sys-bth}+\widehat{H}_{src-sys}+\widehat{H}_{sys-bath-sys}=
\sum_{\omega_{i}\in\mathcal{I}_{sys},\Omega_{j}\in\mathcal{I}_{bth}}\eta_{\omega_{i} \Omega_{j}} \,\,\,\Adj{a}_{\omega_{i}}\widehat{b}_{\Omega_{B}}+\\
&+\sum_{\omega_{i}\in\mathcal{I}_{sys},\Omega_{k}^{'}\in\mathcal{I}_{src}}\xi_{\omega_{i} \Omega_{k}^{'}}\,\,\,\Adj{a}_{\omega_{i}}\widehat{c}_{\Omega_{k}^{'}}+\sum_{\omega_{A_i},\omega_{A_j}\in\mathcal{I}_{sys},\Omega_{k}\in\mathcal{I}_{bth}}\,\,\,\chi_{\omega_{i} \omega_{j} \Omega_{k}}\,\,\Adj{a}_{\omega_{i}}\widehat{a}_{\omega_{j}}\Adj{b}_{\Omega_k}+\mathrm{h.c.}
\end{split}
\end{equation} 
where $\eta_{\omega_i \Omega_j},\xi_{\omega_i \Omega_{k}^{'}},\chi_{\omega_i \omega_j \Omega_k}\in\mathbb{C}$ are the coupling constants describing the linear interactions among the thermal bath modes and the biomolecule modes, the  linear interactions between the external source and the biomolecule, and the  mode-mode interactions among the biomolecule normal modes mediated by the thermal bath, respectively.\\
From these terms it is possible to derive the Fr\"ohlich rate equations, by resorting to time dependent perturbation theory; details are given in Refs. \cite{wu1,wu2} and in the reference book in Ref. \cite{pokorny}.\\
However, the mode coupling term corresponds to a potential energy unbounded  from below and, consequently, this would give rise to dynamical instability of the system and, in the quantum context, to the \textit{absence of a finite energy ground state}. This problem led to strong criticism against the Wu-Austin Hamiltonian model and also against the ensemble of Fr\"ohlich condensation theory \cite{bolterauer}.
This problem can be easily fixed by adding a term with a quartic dependence on the creation/annihilation operators of the form
\begin{equation}
\label{eq:HintQuan_Quartic}
\begin{split}
\widehat{H}_{intQ}=&\sum\limits_{\omega_i,\omega_j,\omega_k,\omega_l\in\mathcal{I}_{sys}}\,\,\Bigr[\kappa_{(1)\omega_i \omega_j \omega_k \omega_l}\,\,\Adj{a}_{\omega_i}\Adj{a}_{\omega_j} \widehat{a}_{\omega_k} \widehat{a}_{\omega_l}+\\
&+\kappa_{(2)\omega_i \omega_j \omega_k \omega_l}\,\,\Adj{a}_{\omega_i}\Adj{a}_{\omega_j} \Adj{a}_{\omega_k} \widehat{a}_{\omega_l}+\kappa_{(3)\omega_i \omega_j \omega_k \omega_l}
\Adj{a}_{\omega_i}\Adj{a}_{\omega_j} \Adj{a}_{\omega_k} \Adj{a}_{\omega_l}\Bigr]+\mathrm{h.c.}
\end{split}
\end{equation}
so that the lower bound to the ground energy level does not go to $-\infty$ for
large values of $N_{\omega_i}$.\\
This quartic interaction stands for a anharmonic interaction among the normal modes of the biomolecule, a broadly studied topic of relevance to energy transport in biomolecules like proteins.\\
 \subsection{Dequantization of Wu-Austin Hamiltonian by Time Dependent Variational Principle (TDVP)}
In this section, a classical Hamiltonian system with its canonical equations is associated
to the quantum system described by the quantum Hamiltonian of eqs. \eqref{eq:HtotQuant}, \eqref{eq:HintQuan_WuAustin} and \eqref{eq:HintQuan_Quartic}. This result is obtained by 
applying the Time Dependent Variational Principle (TDVP) technique  \cite{kramer1980,kramer2008}. The same technique, more deepened from both the mathematical and conceptual points of view, has been proposed as a ``dequantization'' technique in \cite{jauslin} as a kind inverse procedure with respect to the geometrical quantization.
This consists in the evaluation of the time-dependent operator action on coherent states of quantum harmonic oscillators. The scalar parameters describing the coherent states become generalized coordinates of a classical dynamical system whose equations of motion can be derived from a variational principle.\\
In more details, one begins with the ansatz that the wavefunction depends on $N$ parameters
$\{x_i\}_{i=1,...,N}$ 
\begin{equation}
\label{eq:Ansatz_TDVP}
\ket{\psi}=\ket{\psi(x_1,...,x_N)}\,.
\end{equation}
where the parameters $x_i=x_i(t)$ are in general functions of time.
For a quantum system with Hamiltonian $\widehat{H}_{Tot}$ the equations of motion
of the $x_i$ can be derived using the following variational principle (equivalent to
the Least Action principle)
\begin{equation}
\label{eq:VariationalPrinciple_TDVP}
\delta S=0 \qquad \text{with} \qquad S=\int_{0}^{t} L(\psi,\bar{\psi}) \,\, \D t' 
\end{equation}
where $L(\psi,)$ is the Lagrangian associated to the system
\begin{equation}
L(\psi,\bar{\psi})=\dfrac{\imath\hslash}{2}\dfrac{\braket{\psi}{\dot{\psi}}-\braket{\dot{\psi}}{\psi}}{\braket{\psi}{\psi}}-\dfrac{\bra{\psi}\widehat{H}_{Tot}\ket{\psi}}{\braket{\psi}{\psi}} \,\,\,.
\end{equation}
The equations of motions derived from Eq.\eqref{eq:VariationalPrinciple_TDVP}
can be worked out in the framework of classical Hamiltonian dynamics.\\
The classical Hamiltonian is associated with the quantum one by simply taking the expectation value of the Hamiltonian operator $\widehat{H}_{tot}$ over the state $\ket{\psi(x_1,...,x_N)}$, that is

\begin{equation}
H_{Tot}=\bra{\psi(x_1,...,x_N)}\widehat{H}_{Tot}\ket{\psi(x_1,...,x_N)}\,\,.
\end{equation}
The Poisson brackets $\{\cdot,\cdot\}$ depend only on the chosen parametrization
for the wavefunction. Starting from the variables
\begin{equation}
\label{eq:littleW_TDVP}
w_{i}=\imath\hslash\braket{\psi}{\dfrac{\partial \psi}{\partial x_{i}}}=-\imath\hslash\braket{\dfrac{\partial \psi}{\partial x_i}}{\psi}
\end{equation}
one defines the antisymmetric tensor
\begin{equation}
\label{eq:MatrixW_TDVP}
W_{ij}=\dfrac{\partial w_{j}}{\partial x_i}-\dfrac{\partial w_{i}}{\partial x_j}
\end{equation}
so that the equations of motion are implicitly given by
\begin{equation}
\sum_{j=1}^{N}\,W_{ij}\dot{x}_j=\dfrac{\partial H_{Tot}}{\partial x_{i}} \,.
\end{equation}
If the condition $\mathrm{Det} W_{ij}\neq 0$ holds, then the matrix $\mathcal{W}_{ij}=\left(W^{-1}\right)_{ij}$ defines the Poisson brackets for the classical Hamiltonian system
\begin{equation}
\left\{f,g\right\}=\sum_{i,j}^N\,\,\dfrac{\partial f}{\partial x_i}\mathcal{W}_{ij}\dfrac{\partial g}{\partial x_j}\,.
\end{equation}
This formalism can be applied to the quantum system described by the quantum Hamiltonian of Eq.\eqref{eq:HtotQuant} to associate to it a classical Hamiltonian system.
The choice of the parametrization for the wavefunction is quite arbitrary and the TDVP, as any other variational principle, restricts the dynamics to a certain region of the Hilbert space.
Since the Hamiltonian is expressed in terms of creation/annihilation operators of the quantum
harmonic oscillators describing the system, the wavefunction is 
been chosen as a product of the corresponding coherent states. In particular 
\begin{equation}
\label{eq:QuantumState_TotSys}
\ket{\Psi (t)}=\prod_{\omega_i\in\mathcal{I}_{sys},\Omega_j\in\mathcal{I}_{bth},\Omega_{k}^{'}\in\mathcal{I}_{src}}\ket{z_{\omega_{i}}(t)}_{sys}\otimes\ket{z_{\Omega_j}(t)}_{bth}\otimes\ket{z_{\Omega_k^{'}}(t)}_{src}
\end{equation}
where $\ket{z_{A_{i}}(t)}_{sys},\ket{z_{B_{k}}(t)}_{bth},\ket{z_{C_{p}}(t)}_{src}$ are \textit{normalized} coherent states for the normal modes of the main system, of the thermal bath and of the external source, respectively: their general form is given by
\begin{equation}
\label{eq:NormalizedCoherentStates_def}
\ket{z}=\exp\left[-\dfrac{|z|^2}{2}\right]\sum_{k=0}^{+\infty}\dfrac{z^{k}}{\sqrt{k!}}\ket{k}=\exp\left[-\dfrac{|z|^2}{2}\right]\sum_{k=0}^{+\infty}\dfrac{\left(z\widehat{a}^{\dagger}\right)^{k}}{k!}\ket{0} \qquad \text{where}\quad  z=z(t)\in\mathbb{C}\,.
\end{equation}
From the definition of coherent states in Eq.\eqref{eq:NormalizedCoherentStates_def} it follows that the expectation value for the occupation number $n_i$ is given by the squared norm of $z$
\begin{equation}
\left\langle n(t)\right\rangle=\bra{z(t)}\Adj{a}\widehat{a}\ket{(t)}=|z(t)|^2
\end{equation}
so, as we are interested in writing rate equations for these quantities, we parametrize the wavefunction $\psi(t)$ with the set of real parameters $\left\{(n_i,\theta_i)\right\}_{i\in\mathcal{I}_{\mathcal{S}}}$ such that
\begin{equation}
z_i=n_i^{1/2}\exp\left[-\imath\theta_i\right]\quad \Longrightarrow \quad z_{i}^{*}=n_i^{1/2}\exp\left[\imath\theta_i\right] \qquad n_{i}=|z_i|^2\qquad i\in\mathcal{I}_{\mathcal{S}}.
\end{equation}
Using Eq.\eqref{eq:littleW_TDVP} it is possible to derive the Poisson brackets associated with the 
variables $\left\{(n_i,\theta_i)\right\}_{i\in\mathcal{I}_{\mathcal{S}}}$:
\begin{equation}
\begin{split}
&w_{n_i}=\imath\hslash\braket{\left\{(n_j,\theta_j)\right\}_{j\in\mathcal{I}_{\mathcal{S}}}}{\dfrac{\partial}{\partial n_i}\left\{(n_j,\theta_j)\right\}_{j\in\mathcal{I}_{\mathcal{S}}}}=\\
&=\dfrac{\imath\hslash}{\partial n_{i}/\partial |z_i|}\bra{\left\{(n_j,\theta_j)\right\}_{j\in\mathcal{I}_{\mathcal{S}}}}\dfrac{\partial}{\partial |z_i|}\bigotimes_{j\in\mathcal{I}_{\mathcal{S}}}\left(\exp\left[-\dfrac{|z_j|^2}{2}\right]\sum_{k=0}^{+\infty}\dfrac{|z_j|^k \exp\left[-\imath k\theta_{j}\right]\left(\Adj{a}_j\right)^k}{k!}\ket{0}\right)=\\
&=\dfrac{\imath\hslash}{2|z_i|}\Bigr(-|z_i|\braket{\left\{(n_j,\theta_j\right\}_{j\in\mathcal{I}_{\mathcal{S}}}}{\left\{(n_j,\theta_j)\right\}_{j\in\mathcal{I}_{\mathcal{S}}}}+\exp[-\imath\theta_i]\bra{\left\{(n_j,\theta_j)\right\}_{j\in\mathcal{I}_{\mathcal{S}}}}\Adj{a}_{j}\ket{\left\{(n_j,\theta_j)\right\}_{j\in\mathcal{I}_{\mathcal{S}}}}\Bigr)=\\
&=\dfrac{\imath\hslash}{2|z_i|}(-|z_i|+|z_i|\exp[\imath\theta_i]\exp[-\imath\theta_i])=0\\
\end{split}
\end{equation}
\begin{equation}
\begin{split}
&w_{\theta_i}=\imath\hslash\bra{\left\{(n_j,\theta_j)\right\}_{j\in\mathcal{I}_{\mathcal{S}}}}\dfrac{\partial}{\partial\theta_i}\ket{\left\{(n_j,\theta_j)\right\}_{j\in\mathcal{I}_{\mathcal{S}}}}=\\
&=\imath\hslash\bra{\left\{(n_j,\theta_j)\right\}_{j\in\mathcal{I}_{\mathcal{S}}}}\dfrac{\partial}{\partial \theta_i}\bigotimes_{j\in\mathcal{I}_{\mathcal{S}}}\left(\exp\left[-\dfrac{|z_j|^2}{2}\right]\sum_{k=0}^{+\infty}\dfrac{|z_j|^k \exp\left[-\imath k\theta_{j}\right]\left(\Adj{a}_j\right)^k}{k!}\ket{0}\right)=\\
&=\imath(-\imath)\hslash|z_i|\exp[-\imath\theta_i]\bra{\left\{(n_j,\theta_j)\right\}_{j\in\mathcal{I}_{\mathcal{S}}}}\Adj{a}_i\ket{\left\{(n_j,\theta_j)\right\}_{j\in\mathcal{I}_{\mathcal{S}}}}=\hslash|z_i|^2=\hslash n_{i}\,.
\end{split}
\end{equation}
and, consequently, using the definition \eqref{eq:MatrixW_TDVP} the entries of the matrix $W$ are
\begin{align}
&W_{\theta_i \theta_k}=W_{n_i n_k}=0\\
&W_{n_i \theta_k}=-W_{\theta_k n_i}=\dfrac{\partial w_{n_k}}{\partial \theta_i}-\dfrac{\partial w_{\theta_i}}{\partial n_k}=\hslash\delta_{i,k}\,.
\end{align}
and its inverse has the form
\begin{align}
&\mathcal{W}_{\theta_i \theta_k}=\mathcal{W}_{n_i n_k}=0\\
&\mathcal{W}_{\theta_i n_k}=-\mathcal{W}_{n_k \theta_i}=\dfrac{\delta_{i,k}}{\hslash}\,\,.
\end{align}
Consequently it follows that the variables $J_{\omega}=\hslash n_{\omega}$ and $\theta_{\omega}$
are canonically conjugated variables. The classical Hamiltonian $H=H_0+H_{IntWA}+H_{IntQuad}$ for the  variables $\left\{(\theta_{\omega},J_{\omega})\right\}_{\omega\in\mathcal{I}_{\mathcal S}}$ is given by a free classical part
\begin{equation}
\label{eq:ClassH0}
\begin{split}
H_0 =&\bra{\Psi(\theta_{\omega},J_{\omega})}\widehat{H}_{0}\ket{\Psi(\theta_{\omega},J_{\omega})}=\sum_{\omega_i\in\mathcal{I}_{sys}} \omega_i  J_{\omega_i} + \sum_{\Omega_k\in\mathcal{I}_{bth}} \Omega_k J_{\Omega_k}+\\
&+\sum_{\Omega_{p}^{'}\in\in\mathcal{I}_{src} } \Omega'_p J_{\Omega^{'}_p}
\end{split}
\end{equation}
by a semi-classical Wu and Austin interaction part
\begin{equation}
\begin{split}
H_{IntWA}&=\bra{\Psi(\theta_{\omega},J_{\omega})}\widehat{H}_{IntWA}\ket{\Psi(\theta_{\omega},J_{\omega})}=\\
&=\sum\limits_{\omega_i \in \mathcal{I}_{sys}}\sum\limits_{\Omega_k \in \mathcal{I}_{bth}}\dfrac{\left|\eta_{\omega_i \Omega_k}\right|}{\hslash} J_{\omega_i}^{1/2} J_{\Omega_k}^{1/2}\cos\left(\theta_{\omega_i}-\theta_{\Omega_k}+\theta_{\eta_{\omega_i \Omega_k}}\right)+\\
&+\sum\limits_{\omega_i \in \mathcal{I}_{sys}}\sum\limits_{\Omega_p^{'} \in \mathcal{I}_{src}} \dfrac{\left|\xi_{\omega_i \Omega_{p}^{'}}\right|}{\hslash} J_{\omega_i}^{1/2} J_{\Omega_k^{'}}^{1/2}\cos\left(\theta_{\omega_i}-\theta_{\Omega_p^{'}}+\theta_{\xi_{\omega_i \Omega_{p}^{'}}}\right)+\\
&+\sum\limits_{\omega_i,\omega_j\in\mathcal{I}_{sys}}\sum_{\Omega_k\in\mathcal{I}_{bth}}\dfrac{\left|\chi_{\omega_i \omega_j \Omega_k}\right|}{\hslash^{3/2}}J_{\omega_i}^{1/2} J_{\omega_j}^{1/2} J_{\Omega_k}^{1/2}\cos\left(\theta_{\omega_i}-\theta_{\omega_j}+\theta_{\Omega_k}+\theta_{\chi_{\omega_i \omega_j \Omega_k}}\right)
\end{split}
\end{equation}
and by the quartic term
\begin{equation}
\begin{split}
H_{IntQ}&=\bra{\Psi(\theta_{\omega},J_{\omega})}\widehat{H}_{IntQ}\ket{\Psi(\theta_{\omega},J_{\omega})}=
\sum_{\omega_i,\omega_j,\omega_k,\omega_l\in\mathcal{I}_{sys}} J_{\omega_i}^{1/2}J_{\omega_j}^{1/2}J_{\omega_k}^{1/2}J_{\omega_l}^{1/2}\\
&\Biggr[\dfrac{\left|\kappa_{(1)\omega_i \omega_j \omega_k \omega_l}\right|}{\hslash^2}\cos\left(\theta_{\omega_i}+\theta_{\omega_j}-\theta_{\omega_k}-\theta_{\omega_l}+\theta_{\kappa_{(1)\omega_i \omega_j \omega_k \omega_l}}\right)+\\
&\dfrac{\left|\kappa_{(2)\omega_i \omega_j \omega_k \omega_l}\right|}{\hslash^2}\cos\left(\theta_{\omega_i}+\theta_{\omega_j}+\theta_{\omega_k}-\theta_{\omega_l}+\theta_{\kappa_{(2)\omega_i \omega_j \omega_k \omega_l}}\right)+\\
&+\dfrac{\left|\kappa_{(3)\omega_i \omega_j \omega_k \omega_l}\right|}{\hslash^2}\cos\left(\theta_{\omega_i}+\theta_{\omega_j}+\theta_{\omega_k}+\theta_{\omega_l}+\theta_{\kappa_{(3)\omega_i \omega_j \omega_k \omega_l}}\right)\Biggr]
\end{split}
\end{equation}
where each complex coupling constant is given in polar representation.
In what follows, the coupling constants are considered real and rescaled s.t.
\begin{equation}
\begin{split}
&\theta_{\eta_{\omega_i \Omega_k}}=\theta_{\xi_{\omega_i \Omega_{p}^{'}}}=\theta_{\chi_{\omega_i\omega_j \Omega_k}}=\theta_{\kappa_{(1,2,3)\omega_i \omega_j \omega_k \omega_k}}=0 \\
&\dfrac{\left|\eta_{\omega_i \Omega_k}\right|}{\hslash}\rightarrow \eta_{\omega_i \Omega_K}, \quad \dfrac{|\xi_{\omega_i\Omega_p^{'}}|}{\hslash}\rightarrow \xi_{\omega_i \Omega_{p}^{'}},  \quad \dfrac{\left|\chi_{\omega_i \omega_j \Omega_k}\right|}{\hslash^{3/2}}\rightarrow \chi_{\omega_i \omega_j \Omega_k}, \quad \dfrac{\left|\kappa_{(1,2,3)\omega_i \omega_j \omega_k \omega_l}\right|}{\hslash^2}\rightarrow \kappa_{(1,2,3)\omega_i \omega_j \omega_k \omega_l}\\
\end{split} 
\end{equation}
with these choices, the total Hamiltonian of the system reads
\begin{equation}
\label{eq:HamTot_Classic}
\begin{split}
&H_{Tot}(\left\{(J_{\omega_i},\theta_{\omega_i})\right\}_{\omega_i\in\mathcal{I}_{\mathcal{S}}})=\sum_{\omega_i\in\mathcal{I}_{sys}}\,\omega_{i} J_{i}+\sum_{\Omega_j\in\mathcal{I}_{bth}}\,\Omega_{j} J_{\Omega_j}+\sum_{\Omega_k^{'}\in\mathcal{I}_{src}} \,\Omega_k^{'} J_{{\Omega_k}^{'}}+\\
&+\sum_{\omega_i\in\mathcal{I}_{sys}}\sum_{\Omega_j\in\mathcal{I}_{bth}}\,\eta_{\omega_i \Omega_j}J_{\omega_i}^{1/2}J_{\Omega_j}^{1/2}\,\cos\left(\theta_{\omega_i}-\theta_{\Omega_j}\right)+\sum_{\omega_i\in\mathcal{I}_{sys}}\sum_{\Omega_k^{'}\in\mathcal{I}_{src}}\,\xi_{\omega_i \Omega_k^{'}}J_{\omega_i}^{1/2}J_{\Omega_k^{'}}^{1/2}\cos\left(\theta_{\omega_i}-\theta_{\Omega_k^{'}}\right)+\\
&\sum\limits_{\omega_i,\omega_j\in\mathcal{I}_{sys}}\sum_{\Omega_k\in\mathcal{I}_{bth}}\chi_{\omega_i \omega_j \Omega_k}\,\, J_{\omega_i}^{1/2} J_{\omega_j}^{1/2} J_{\Omega_k}^{1/2}\cos\left(\theta_{\omega_i}-\theta_{\omega_j}+\theta_{\Omega_k}\right)+\\
&+\sum_{\omega_i,\omega_j,\omega_k,\omega_l\in\mathcal{I}_{sys}} J_{\omega_i}^{1/2}J_{\omega_j}^{1/2}J_{\omega_k}^{1/2}J_{\omega_l}^{1/2} \Biggr[\kappa_{(1)\omega_i \omega_j \omega_k \omega_l}\cos\left(\theta_{\omega_i}+\theta_{\omega_j}-\theta_{\omega_k}-\theta_{\omega_l}\right)+\\
&\kappa_{(2)\omega_i \omega_j \omega_k \omega_l}\cos\left(\theta_{\omega_i}+\theta_{\omega_j}+\theta_{\omega_k}-\theta_{\omega_l}\right)+\kappa_{(3)\omega_i \omega_j \omega_k \omega_l}\cos\left(\theta_{\omega_i}+\theta_{\omega_j}+\theta_{\omega_k}+\theta_{\omega_l}\right)\Biggr]\,\,.
\end{split}
\end{equation}
In order to derive Fr\"ohlich-like rate equations, the dynamics of the action variables $J_{\omega_i}$ of the system  has to be studied. We could choose to investigate the dynamics of the system by letting observable quantities evolve in time (according to Hamilton's equations of motion) and then performing time averaging, and averaging on different initial conditions compatible with the assumption that the two subsystems $\mathcal{I}_{bth}$ and $\mathcal{I}_{src}$ are two thermal baths with different temperatures. However, this method has some disadvantages: the integration should be done numerically because of the presence of non-linear interactions terms in the Hamiltonian, and a very large number of degrees of freedom would be necessary to adequately simulate the dynamics of the thermal baths. 
Moreover, long integration times would be necessary to  attain the convergence of time averages of the observables, and, finally, this computational effort would not be worth as it would provide a redundant information on the dynamics.
Another way to derive rate equations for the $J_{\omega_i}$ consists in a statistical 
approach: the relevant dynamical variable to consider is the phase space distribution function $\rho(\left\{(J_{\omega},\theta_{\omega})\right\}_{\omega\in\mathcal{I}_{\mathcal{S}}},t)$ so that the rate equations are written for the \textit{statistical averages of actions variables}
\begin{equation}
\left\langle J_{\omega_i}(t)\right\rangle=\int\,\,J_{\omega_i}\rho(\left\{\left(J_{\omega},\theta_{\omega}\right)\right\}_{\omega\in\mathcal{I}_{\mathcal{S}}},t)\prod\limits_{\omega\in\mathcal{I}_{\mathcal{S}}}\D J_{\omega_i}\D\theta_{\omega_i}\,\,\,.
\end{equation}
In the following Section  classical Fr\"ohlich-like rate equations are derived by resorting to the time evolution of the distribution function $\rho(\left\{(J_{\omega},\theta_{\omega})\right\}_{\omega\in\mathcal{I}_{\mathcal{S}}},t)$ satisfying the Liouville equation. 
 
 \subsection{Derivation of classical rate equations using the Koopman-Von Neumann formalism}
 Let $\rho(\left\{(J_{\omega},\theta_{\omega})\right\}_{\omega\in\mathcal{I}_{\mathcal{S}}};t)$ be a probability density function for the whole system described by the Hamiltonian in Eq.\eqref{eq:HamTot_Classic};
according to the Liouville Theorem the evolution of $\rho$ associated with this Hamiltonian is given by
\begin{equation}
\label{eq:LiouvilleTh_BC}
\dfrac{\partial \rho}{\partial t}=\left\{H,\rho\right\}=-\imath\mathcal{L}_{H}(\rho)
\end{equation}
where $\imath^2=-1$ and $\left\{\cdot,\cdot\right\}$ are the canonical Poisson brackets
\begin{equation}
\begin{split}
\left\{f,g\right\}=&\sum_{\omega_i}\left(\dfrac{\partial f}{\partial J_{\omega_i}}\dfrac{\partial g}{\partial \theta_{\omega_i}}-\dfrac{\partial g}{\partial J_{\omega_i}}\dfrac{\partial f}{\partial \theta_{\omega_i}}\right)+\sum_{\Omega_j}\left(\dfrac{\partial f}{\partial J_{\Omega_j}}\dfrac{\partial g}{\partial \theta_{\Omega_j}}-\dfrac{\partial g}{\partial J_{\Omega_j}}\dfrac{\partial f}{\partial \theta_{\Omega_j}}\right)+\\
&+\sum_{\Omega_k^{'}}\left(\dfrac{\partial f}{\partial J_{\Omega_k^{'}}}\dfrac{\partial g}{\partial \theta_{\Omega_k^{'}}}-\dfrac{\partial g}{\partial J_{\Omega_k^{'}}}\dfrac{\partial f}{\partial \theta_{\Omega_k^{'}}}\right)\,.
\end{split}
\end{equation}
and $\mathcal{L}_{H}(\cdot)=\imath\left\{H,\cdot\right\}$ is the \textit{Liouville operator}
acting on functions defined on the phase space of the system. An interesting method to study and solve Liouville equations relies on the Koopman-Von Neumann (KvN) formalism developed in the 30's: a formal
analogy among Liouville and Schr\"odinger equation is established so that also classical mechanics can be formulated in the framework of a Hilbert space of square integrable functions.\\
In our case the Hilbert space of complex square integrable functions in phase space is 
$L^{2}(\Lambda_{\{(J_{\omega},\theta_{\omega})\}_{\omega\in\mathcal{I}_{\mathcal{S}}}})$ with the inner product defined by
\begin{equation}
\braket{f}{g}=\int_{\lambda_{\{(J,\alpha)\}_{\omega\in\mathcal{I}_{\mathcal{S}}}}}\,f^{*} g \,\,\prod\limits_{\omega\in\mathcal{I}_{\mathcal{S}}}\D J_{\omega}\,\D\theta_{\omega}=\prod\limits_{\omega\in\mathcal{I}_{\mathcal{S}}}\int_{0}^{2\pi}\D\theta_{\omega}\int_{0}^{+\infty}\D J_{\omega}\,\,f^{*} g 
\end{equation}
with $f,g\in L^{2}(\Lambda_{\{(J_{\omega},\theta_{\omega})\}_{\omega\in\mathcal{I}_{\mathcal{S}}}})$.
On this space we can define the action of the Liouville operator
\begin{equation}
\widehat{\mathcal{L}}_{H}\ket{f}=\mathcal{L}_{H}(f)\,\,.
\end{equation}
and consider the domain $\mathcal{D}_{\widehat{\mathcal{L}}_{H}}\subseteq L^{2}(\Lambda_{\{(J_{\omega},\theta_{\omega})\}_{\omega\in\mathcal{I}_{\mathcal{S}}}})$ where the Liouville operator is self-adjoint, namely, $\Adj{\mathcal{L}}_{H}=\widehat{\mathcal{L}_{H}}$ and $\mathcal{D}_{\widehat{\mathcal{L}}_{H}}=\mathcal{D}_{\Adj{\mathcal{L}}_{H}}$.\\
Let $\psi(\{(J,\theta)\}_{\mathcal{I}_\mathcal{S}};t)\in\mathcal{D}_{\widehat{\mathcal{L}}_{H}}$
be a normalized time-dependent function \cite{nota} such that
\begin{equation}
\imath\dfrac{\partial \psi}{\partial t}(\mathbf{J},\boldsymbol{\theta};t)=\widehat{\mathcal{L}}_{H}\psi (\mathbf{J},\boldsymbol{\theta};t)
\end{equation}
then it can be proved that $\rho=\|\psi\|_{L^{2}(\Lambda_{(J,\theta)})}=\psi^{*}\psi$ is a normalized function for which \eqref{eq:LiouvilleTh_BC} holds. Moreover as $\widehat{\mathcal{L}}_{H}$ is a self-adjoint operator, it represents the unitary time evolution of the ``wave function'' as
\begin{equation}
\psi(\{(J,\theta)\}_{\mathcal{I}_\mathcal{S}};t)=\exp\left[-\imath t \widehat{\mathcal{L}}_{H}\right]\psi(\{(J,\theta)\}_{\mathcal{I}_\mathcal{S}};0)
\end{equation}
in analogy with quantum mechanics.
With this formalism the rate equations  for the average values of the actions (which are the counterpart of quantum occupation numbers) associated with the normal modes $\omega_i$ of the main system  are given by
\begin{equation}
\label{eq:meanofaction_2}
\dfrac{\D}{\D t}\left\langle J_{\omega_i} \right\rangle_{t}=\dfrac{\D}{\D t}\bra{\psi (t)}\widehat{\mathcal{M}}_{ J_{\omega_i}}\ket{\psi(t)}=\imath\bra{\psi(t)}\left[\widehat{\mathcal{L}}_{H},\widehat{\mathcal{M}}_{ J_{\omega_i}}\right]\ket{\psi(t)}
\end{equation}
where $\widehat{\mathcal{M}}_{J_{\omega_i}}$ is a multiplicative operator acting on $L^{2}(\Lambda_{\{(J_{\omega},\theta_{\omega})\}_{\omega\in\mathcal{I}_{\mathcal{S}}}})$ as follows
\begin{equation}
\widehat{\mathcal{M}}_{J_{\omega_i}}\ket{\psi}=\ket{J_{\omega_i}\psi}\,.
\end{equation}
The Liouville operator can be decomposed as
\begin{equation}
\widehat{\mathcal{L}_{H}}=\widehat{\mathcal{L}}_{H_{0}}+\widehat{\mathcal{L}}_{H_{Int}}(t)\ ,
\end{equation}
and since the eigenfunctions of the operator $\widehat{\mathcal{L}}_{H_{0}}$ are known, the action of 
the operator $\widehat{\mathcal{L}}_{H_{Int}}$ can be treated as a time dependent perturbation, which is adiabatically turned on and off from $t_0=0$ to $t=+\infty$.
Then a classical analogous of the interaction representation formalism is used.  Thus, if $\ket{\psi (t)}_{S}$ is the ``wave function'' 
in the Schr\"odinger representation, then in interaction representation $\ket{\psi (t)}_{I}$
reads
\begin{equation}
\ket{\psi (t)}_{I}=\exp\left[\imath t\widehat{\mathcal{L}}_{H_{0}}\right]\ket{\psi (t)}_{S}
\end{equation}
and, given a generic operator $\widehat{A}_{S}$ in the Schr\"odinger picture, its expression in 
the interaction picture $\widehat{A}_{I}$ ris 
\begin{equation}
\label{eq:IP_operatorTimeEvolution}
\widehat{A}_{I}(t)=\exp\left[\imath t\widehat{\mathcal{L}}_{H_{0}}\right]\widehat{A}_{S}\exp\left[-\imath t\widehat{\mathcal{L}}_{H_{0}}\right]\,.
\end{equation}
With this formalism the time evolution of $\ket{\psi (t)}_{I}$ can be written through the unitary evolution operator $\widehat{U}(t;t_{0})$ satisfying
\begin{align}
&\ket{\psi (t)}_{I}=\widehat{U}(t;t_{0})\ket{\psi (t_0)}_{I}\\
\label{eq:UnitaryEvo_InteracPict}
&\imath\dfrac{\partial U(t;t_{0})}{\partial t}=\widehat{\mathcal{L}}_{H_{int}}(t)\widehat{U}(t;t_{0})
\end{align}
and the formal solution of \eqref{eq:UnitaryEvo_InteracPict} is given by
\begin{equation}
\label{eq:UnitaryEvo_FormalSolution}
U(t;t_{0})=\mathbb{I}-\imath\int_{t_ {0}}^{t}\,\,\widehat{\mathcal{L}}_{H_{int}}(t')\widehat{U}(t';t_{0}) \,\,\D t' \,.
\end{equation}
At  first order in $\widehat{\mathcal{L}}_{H_{int}}(t')$, the unitary evolution operator $\widehat{U}(t;t_{0})$ in the right hand side of eq.\label{eq:UnitaryEvo_FormalSolution} is substituted by
the identity operator meaning that the state $\ket{\psi(t_0)}_{I}$, if the perturbation is turned on at $t_0$, at the lowest order can be approximated 
by $\ket{\psi (t_0)}_{I}\simeq \ket{\psi(t_0)}_{S}$ and assumed to be coincident with Schr\"odinger picture (i.e. $\ket{\psi(0)}=\ket{\psi_{0}}$), so that
\begin{equation}
\ket{\psi (t)}_{I}\approx\left(\widehat{\mathbb{I}}-\imath\int_{0}^{t}\,\,\widehat{\mathcal{L}}_{H_{int}}(t')\D t'\right)\ket{\psi_{0}}_{I}
\end{equation}
and, as $\widehat{\mathcal{L}}_{H_{int}}(t')$ is self-adjoint, the time evolution for the "bra"  has the form
\begin{equation}
\bra{\psi(t)}_{I}\approx\bra{\psi_{0}}_{I}\left(\widehat{\mathbb{I}}+\imath\int_{0}^{t}\,\,\widehat{\mathcal{L}}_{H_{int}}(t')\D t'\right)\,.
\end{equation}
The time derivative of the multiplicative operator $\widehat{\mathcal{M}}_{J_{\omega_i}}$ in interaction picture is derived according to Eq.\eqref{eq:IP_operatorTimeEvolution}
\begin{equation}
\left(\dot{\widehat{\mathcal{M}}}_{J_{\omega_i}}\right)_{I}=\imath\left(\left[\widehat{\mathcal{L}}_{H},\widehat{\mathcal{M}}_{ J_{\omega_i}}\right]\right)_{I}=\imath\exp\left[\imath t\widehat{\mathcal{L}}_{H_{0}}\right]\left[\widehat{\mathcal{L}}_{H},\widehat{\mathcal{M}}_{ J_{\omega_i}}\right]\exp\left[-\imath t\widehat{\mathcal{L}}_{H_{0}}\right]\,.
\end{equation}
This means that the average \eqref{eq:meanofaction_2} can be entirely rewritten using the interaction picture as
\begin{equation}
\label{eq:rateEquat_KoopVNeu}
\begin{split}
&\dfrac{\D}{\D t}\left\langle J_{\omega_i}\right\rangle_{t}=
_I\bra{\psi(t)}\left(\dot{\widehat{\mathcal{M}}}_{J_{\omega_i}}\right)_{I}\ket{\psi(t)}_{I}=\imath_I\bra{\psi(t)}\left(\left[\widehat{\mathcal{L}}_{H},\widehat{\mathcal{M}}_{ J_{\omega_i}}\right](t)\right)_{I}\ket{\psi(t)}_{I}=\\
&\approx\imath\bra{\psi_{0}}\left(\widehat{\mathbb{I}}+\imath\int_{-\infty}^{t}\widehat{\mathcal{L}}_{H_{int}}(t')\,\D t'\right)\left(\left[\widehat{\mathcal{L}}_{H},\widehat{\mathcal{M}}_{ J_{\omega_i}}\right](t)\right)_{I}\left(\widehat{\mathbb{I}}-\imath\int_{0}^{t}\widehat{\mathcal{L}}_{H_{int}}(t')\,\D t'\right)\ket{\psi_{0}}=\\
&=\imath\bra{\psi_{0}}\left(\left[\widehat{\mathcal{L}}_{H},\widehat{\mathcal{M}}_{ J_{\omega_i}}\right](t)\right)_{I}\ket{\psi_0}+\int_{0}^{t}\,\,\bra{\psi_{0}}\left[\left(\left[\widehat{\mathcal{L}}_{H_{int}},\widehat{\mathcal{M}}_{J_{\omega_i}}\right](t)\right)_{I},\left(\widehat{\mathcal{L}}_{H_{int}}(t')\right)_{I}\right]\ket{\psi_0} \D t'
\end{split}
\end{equation}
After lengthy computations (see SM for the details) one finally arrives at the following rate equations for the expectation values of the actions (that is, the amplitudes of vibrational modes) which are worked out as follows
\begin{equation}
\label{eq:ClassicalFrohlichRateEQs}
\begin{split}
&\dfrac{\D \langle J_{\omega_i} \rangle}{\D t}= s_{\omega_i}+b_{\omega_i}
\left[\dfrac{k_B T_B}{\omega_i}-\left\langle J_{\omega_i}\right\rangle\right]
+\sum_{\omega_j\in\mathcal{I}_{sys} \atop \omega_j\neq\omega_i} c_{\omega_i\omega_j}\Biggr[\Biggr(\left\langle J_{\omega_j}\right\rangle-\left\langle J_{\omega_i}\right\rangle\Biggr)+\dfrac{\omega_j -\omega_i}{k_{B} T_{B}}\left\langle J_{\omega_i}\right\rangle\left\langle J_{\omega_j}\right\rangle\Biggr]+\\
&+\sum_{\omega_j,\omega_k,\omega_l \in \mathcal{I}_{sys}\atop\omega_i+\omega_j-\omega_k-\omega_l=0}\dfrac{16\pi \kappa_{(1)\omega_i\omega_j\omega_k\omega_l}^2}{\delta\omega_{sys}}\left\langle J_{\omega_l}\right\rangle \Biggr(\left\langle J_{\omega_j}\right\rangle \left\langle J_{\omega_k}\right\rangle +\left\langle J_{\omega_i}\right\rangle \left\langle J_{\omega_k}\right \rangle-2 \left\langle J_{\omega_i}\right\rangle \left\langle J_{\omega_j}\right\rangle\Biggr)\\
&+\dfrac{3 \pi }{\delta\omega_{sys}}\Biggr[\sum_{\omega_{j}\omega_{k}\omega_{l}\in \mathcal{I}_{sys} \atop \omega_{i}+\omega_{j}+\omega_{k}-\omega_{l}=0} 3 \kappa_{(2)\omega_i\omega_j\omega_k\omega_l}^2 \left\langle J_{\omega_l}\right\rangle \Biggr(\left\langle J_{\omega_j}\right\rangle \left\langle J_{\omega_k}\right\rangle+2\left\langle J_{\omega_i}\right\rangle \left\langle J_{\omega_k}\right\rangle-\left\langle J_{\omega_i}\right\rangle \left\langle J_{\omega_j}\right\rangle\Biggr)+\\
&+\sum_{\omega_{j}\omega_{k}\omega_{l}\in\mathcal{I}_{sys} \atop \omega_{i}-\omega_{j}-\omega_{k}-\omega_{l}=0}\,\kappa_{(2)\omega_i\omega_j\omega_k\omega_l}^2 \left\langle J_{\omega_l}\right\rangle \left\langle J_{\omega_k} \right\rangle \Biggr(\left\langle J_{\omega_j}\right\rangle-3\left\langle J_{\omega_i}\right\rangle\Biggr)\Biggr].
\end{split}
\end{equation}

\subsection{Properties of the classical rate equations}
The rate equations \eqref{eq:ClassicalFrohlichRateEQs} that have been derived by solving the Liouville equation with the Hamiltonian \eqref{eq:HamTot_Classic} display a functionally similar structure to the original Fr\"ohlich's equations \cite{Frohlich:1968}, apart from the additional quartic terms. 
As in the original formulation given by Fr\"ohlich, the condensation phenomenon is found by considering the stationary solutions of  the rate equations. It is convenient  to rewrite  Eqs.\eqref{eq:ClassicalFrohlichRateEQs} in non-dimensional form after introducing the following variables
\begin{equation}
\begin{split}
& \tau=t\omega_{0} \qquad \text{with} \qquad \omega_{0}=\min_{\omega\in\mathcal{I}_{sys}}\omega \qquad y_{\omega_i}=\dfrac{\omega_{i}\left\langle J_{\omega_{i}}\right\rangle}{k_{\text{B}}T_{B}}\qquad \alpha_{\omega_i}=\dfrac{\omega_i}{\omega_0}\\
& S_{\omega_i}=\alpha_{\omega_i}\dfrac{s_{\omega_i}}{k_{\text{B}}T_B} \qquad B_{\omega_i}=\dfrac{b_{\omega_i}}{\omega_0} \qquad C_{\omega_{i}\omega}=\dfrac{c_{\omega_{i}\omega}}{\omega_0}\qquad \Upsilon_{(1)\omega_i\omega_j\omega_k\omega_l}=\dfrac{16 \pi\kappa_{(1)\omega_i\omega_j\omega_k\omega_l}(k_{\text{B}}T_B)^2}{\delta\omega_{sys} \omega_{0}^3}\\
& \Upsilon_{(2)\omega_i\omega_j\omega_k\omega_l}=\dfrac{3\pi\kappa_{(2)\omega_i\omega_j\omega_k\omega_l}(k_{\text{B}}T_B)^2}{\delta\omega_{sys} \omega_{0}^3}\\
\end{split}
\end{equation}
so that Eqs.\eqref{eq:ClassicalFrohlichRateEQs} read
\begin{equation}
\label{eq:NonDim_ClassicalFrohlEqs}
\begin{split}
&\dot{y}_{\omega_i}=\dfrac{\D y_{\omega_i}}{\D t}=S_{\omega_i}+B_{\omega_i}\left(1-y_{\omega_i}\right)+\sum_{\omega_j\in\mathcal{I}_{sys} \atop \omega_j\neq\omega_i}C_{\omega_{i}\omega_j}\biggr[
\left(\dfrac{\alpha_{\omega_i}}{\alpha_{\omega_j}}y_{\omega_j}-y_{\omega_{i}}\right)+\dfrac{(\alpha_{\omega_j}-\alpha_{\omega_i})}{\alpha_{\omega_j}}y_{\omega_i}y_{\omega_j}\biggr]+\\
&+\sum_{\omega_j,\omega_k,\omega_l \in \mathcal{I}_{sys}\atop\omega_i+\omega_j-\omega_k-\omega_l=0}\Upsilon_{(1)\omega_i\omega_j\omega_k\omega_l}\dfrac{y_{\omega_l}}{\alpha_{\omega_l}} \Biggr(
\dfrac{\alpha_{\omega_i}}{\alpha_{\omega_j}\alpha_{\omega_k}}y_{\omega_j}y_{\omega_k}
 +\dfrac{\alpha_i}{\alpha_k}y_{\omega_i}y_{\omega_k}-2\dfrac{\alpha_{\omega_i}}{\alpha_{\omega_j}}y_{\omega_i}y_{\omega_j}\Biggr)\\
&+\Biggr[\sum_{\omega_{j}\omega_{k}\omega_{l}\in \mathcal{I}_{sys} \atop \omega_{i}+\omega_{j}+\omega_{k}-\omega_{l}=0} 3 \Upsilon_{(2)\omega_i\omega_j\omega_k\omega_l} \dfrac{y_{\omega_l}}{\alpha_{\omega_l}}\Biggr(\dfrac{\alpha_{\omega_i}}{\alpha_{\omega_j}\alpha_{\omega_k}}y_{\omega_j}
y_{\omega_k}+2\dfrac{1}{\alpha_{\omega_k}}y_{\omega_i}y_{\omega_k}-\dfrac{1}{\alpha_{\omega_j}} y_{\omega_i}y_{\omega_j}\Biggr)+\\
&+\sum_{\omega_{j}\omega_{k}\omega_{l}\in\mathcal{I}_{sys} \atop \omega_{i}-\omega_{j}-\omega_{k}-\omega_{l}=0}\,\Upsilon_{(2)\omega_i\omega_j\omega_k\omega_l} \dfrac{y_{\omega_k}y_{\omega_l}}{\alpha_{\omega_k}\alpha_{\omega_l}}\Biggr(\dfrac{\alpha_{\omega_i}}{\alpha_{\omega_j}}y_{\omega_j}-3y_{\omega_i}\Biggr)\Biggr]\qquad\qquad \omega_{i}\in\mathcal{I}_{\omega_i}\,.
\end{split}
\end{equation}
By inspection, one can notice the following properties of the equations above 
\begin{itemize}
\item if the  normal modes thermalize at the heat bath temperature $T_B$, so that
$\left\langle J_{\omega_i} \right\rangle=k_{\text{B}}T_B/\omega_i$, it follows
that the variables $y_{\omega_i}$ are equal and take the value $y_{\omega_i}=1$;

\item by switching-off the external source of energy, i.e. putting $S_{\omega_i}=0$,
the  thermal solution above, that is $y_{\omega_i}=1$ for all the $\omega_i$, is a stationary solution of the system, namely
$\dot{y}_{\omega_i}=0$. 
\end{itemize}
In order to understand whether the phonon condensation phenomenon exists also in the above defined classical framework, one has to work out non trivial out of equilibrium stationary states of the model equations. Unfortunately, doing this analytically for a system with a large number of nonlinear equations is hopeless, therefore one has to resort to numerical integration of the dynamical equations \eqref{eq:NonDim_ClassicalFrohlEqs}.  
A difficulty of the numerical approach is to provide \textit{a priori} estimates of the coupling constants for a real biomolecule. To overcome this problem we borrowed from Ref.\cite{piazza} the values of physically realistic nonlinear coupling parameters. Moreover, the non-dimensional form of the equations partially simplifies the problem as only the ratios among the coefficients has to be known. On the other hand, since we are interested only in a qualitative study of purely theoretical kind, some simplifying assumptions are in order.
Thus, the coupling constants appearing in Eqs.\eqref{eq:NonDim_ClassicalFrohlEqs} are supposed to be independent of the frequency, in particular, also the condition $S_{\omega_i}=S$ is assumed (the energy injection rate is the same for all the normal modes). Then we have performed the numerical integration of the dynamical equations by changing the total number $N+1$ of oscillators (normal modes) in a fixed range ($1\leq\omega_i\leq 2$) of equally spaced frequencies given (in arbitrary units) by  $\omega_{n}=1+n/N$ with $n=0,1,2,..,N$. This choice models an increasing density of modes in a given fixed frequency interval.
The classical energy condensation phenomenon,  numerically found and described below,
is characterised by a strong deviation from energy equipartition among  the normal modes  in a stationary state of the dynamical system \eqref{eq:ClassicalFrohlichRateEQs} (and, equivalently, of the system \eqref{eq:NonDim_ClassicalFrohlEqs} ):   as the energy injection rate increases, the system undergoes a major change in the energy distribution among its normal modes resulting in a more ``organized'' phase (the energy is mainly channeled into the lowest frequency mode).  
Such a condensation phenomenon can be viewed as associated with the breaking of the permutation symmetry for the set of variables $\left\{y_{\omega_0},...,y_{\omega_{N}}\right\}$ in the stationary state, where $y_{\omega_i}$ is the energy content in the mode of frequency $\omega_i$ in $k_{\text{B}}T_B$ units. It is clear that such a symmetry holds at energy equipartition. In analogy with the characterization of a symmetry breaking in the classical theory of equilibrium phase transitions, we have to define some (possibly scalar) indicators to detect the permutation symmetry breaking. In order to detect energy equipartition, and its violation in presence of the condensation phenomenon, we borrow and adapt  from Ref.\cite{reimers} a condensation index  $\mathcal{E}_y$ which reads 
\begin{equation}
\label{entropy}
\mathcal{E}_y=  \dfrac{y_{\omega_0} - {\widetilde{y}_{\omega_0}}}{\sum_{i=0}^Ny_{\omega_i}}
\end{equation}
where $y_{\omega_0}$ is the energy content of the fundamental mode and  ${\widetilde{y}_{\omega_0}} = 1+ S/B$ is the energy content of the fundamental mode given by the exact solution of Eqs.\eqref{eq:NonDim_ClassicalFrohlEqs} in the absence of all the nonlinear couplings (mode-mode and system-bath). Here we replace the occupation numbers of the modes entering the condensation index of Ref.\cite{reimers} with the mode energy contents. $\mathcal{E}_y =0$ corresponds to energy equipartition among the normal modes, and  
$\mathcal{E}_y = 1$ corresponds to full concentration of the energy into the lowest frequency mode. In order to easily compare the results obtained at different number $N$ of modes, the energy fractions $y_{\omega_i}$ in each normal mode are normalised to give the variables  $p_i=y_{\omega_i}/\sum_{\omega_{i}\in\mathcal{I}_{sys}}y_{\omega_i}$ .

Another possible choice for the indicator detecting the permutation symmetry breaking is any of the quantities $y_{\omega_{i}}/y_{\omega_{0}}=p_i/p_0$ with $i\neq 0$: these are expected to attain their maximum value equal to one at equipartition and to take values between zero and one for a Fr\"{o}hlich-like condensation phenomenon. 
Numerical simulations have been performed in order to 
characterize this major change of the stationary states of the dynamical system \eqref{eq:NonDim_ClassicalFrohlEqs}, that is, to find the asymptotic stationary solutions of the variables $y_{\omega_i}(t)$ which correspond to the fixed points $\dot{y}_{\omega_i}=0$ of the system. As a reasonable initial condition,  at $t=0$ the system 
is taken at thermal equilibrium with the heath bath, that is, all the $y_{\omega_i}(0)$ are set equal to $1$.

\subsection{Results of numerical simulations}
\begin{figure}[h!]
 \centering
 \includegraphics[scale=0.3,keepaspectratio=true]{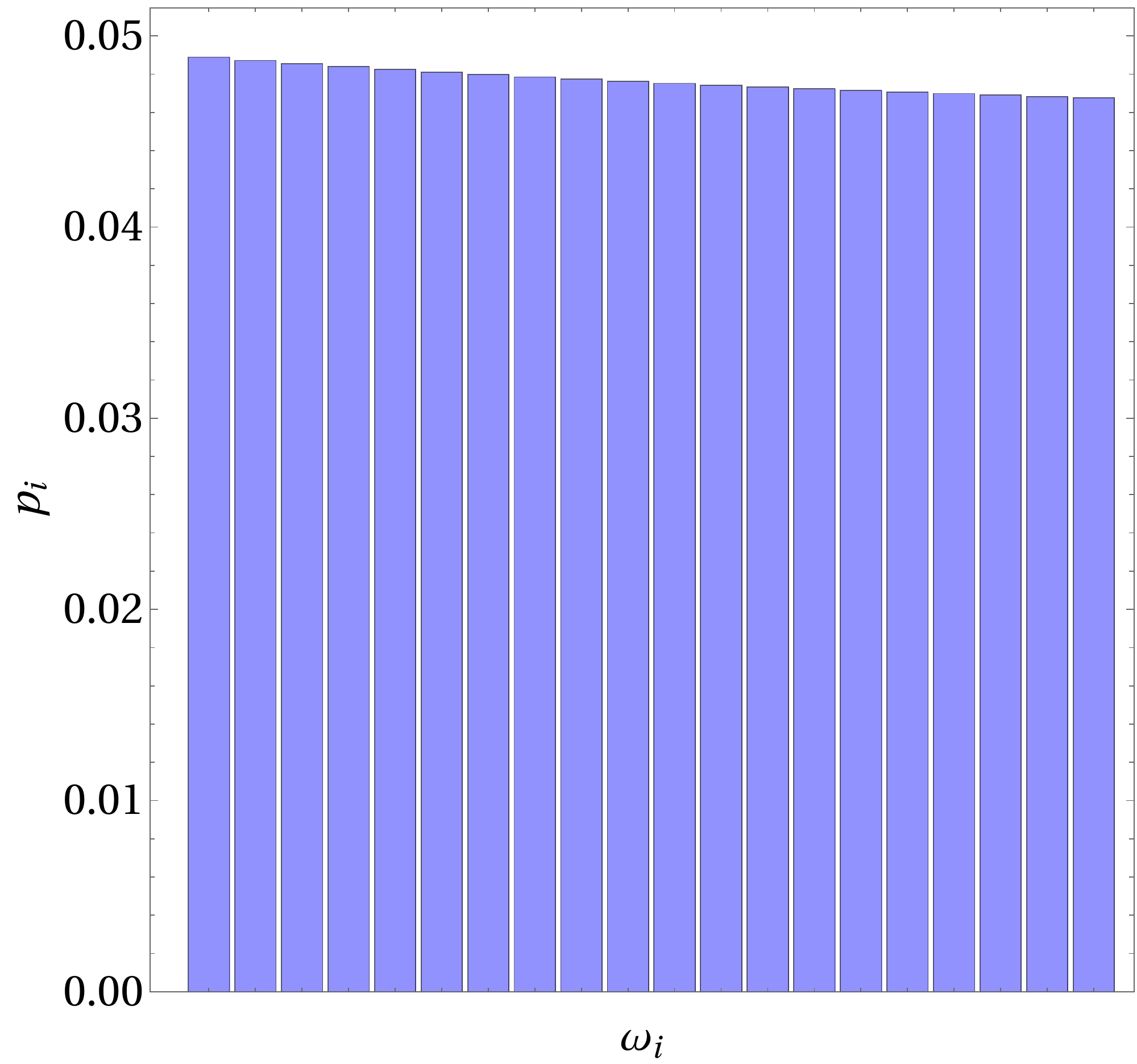}
  \includegraphics[scale=0.3,keepaspectratio=true]{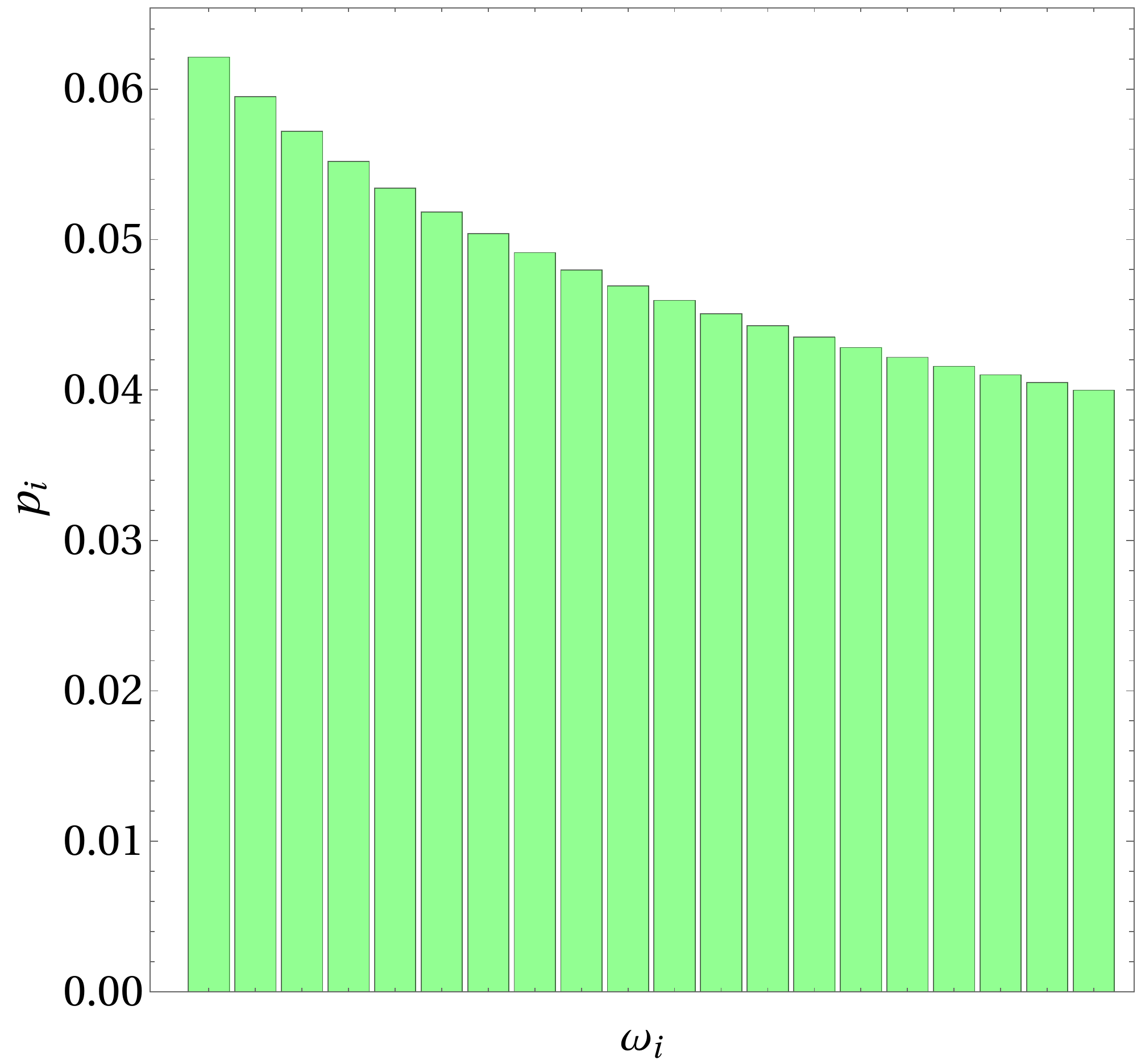}
    \includegraphics[scale=0.3,keepaspectratio=true]{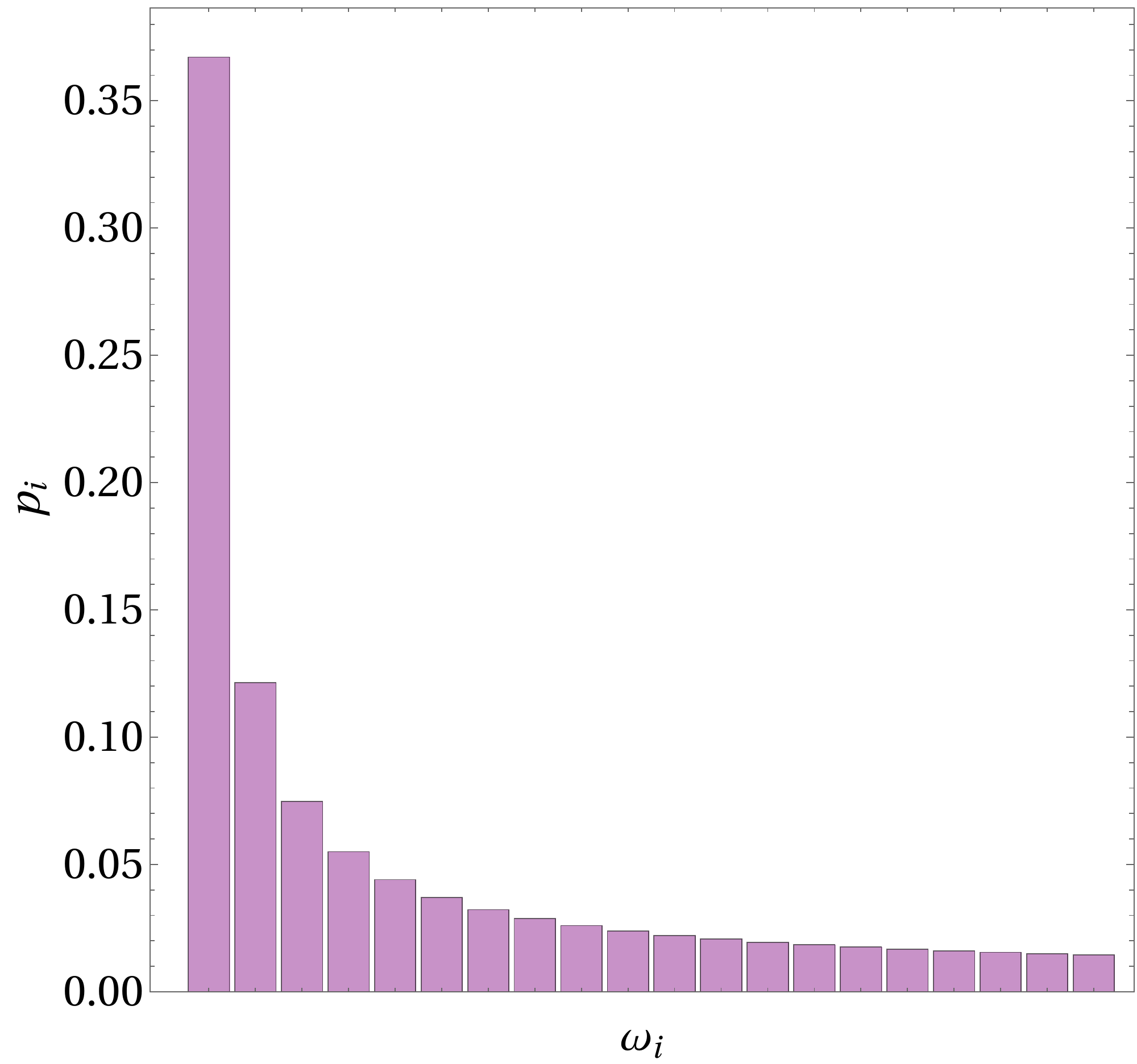}
\includegraphics[scale=0.3,keepaspectratio=true]{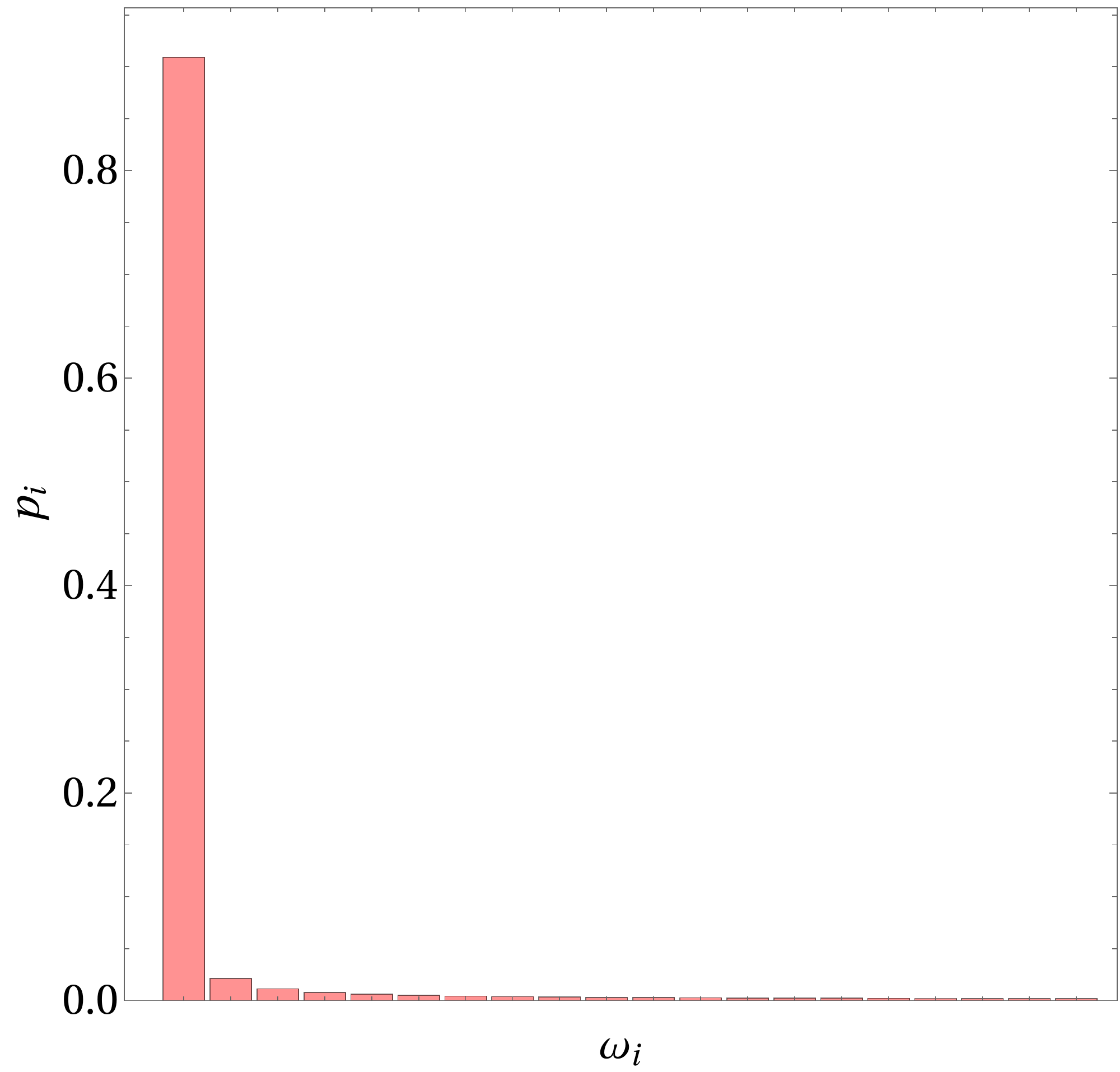}
 \caption{Classical Fr\"ohlich-like condensation. Normalized fractions of energy $p_i$ in normal modes vs. the mode frequencies of the system $\mathcal{I}_{sys}$ for a fixed number of modes: $N=20$. The histograms display increasing deviations from energy equipartition among the normal modes as the energy input rate $S$ increases. The spectra correspond to:  $S=0.1$ (blue), $S=1$ (green), $S=10$ (purple), $S=100$ (pink). Equipartition would correspond to equal heights of the bars. Note the great difference among the energy fraction in  the lowest normal mode and in the higher frequency modes. }
\label{fig:DiagrClasCondens_Sys2}
\end{figure}

\begin{figure}[h!]
\centering
\includegraphics[scale=0.33,keepaspectratio=true]{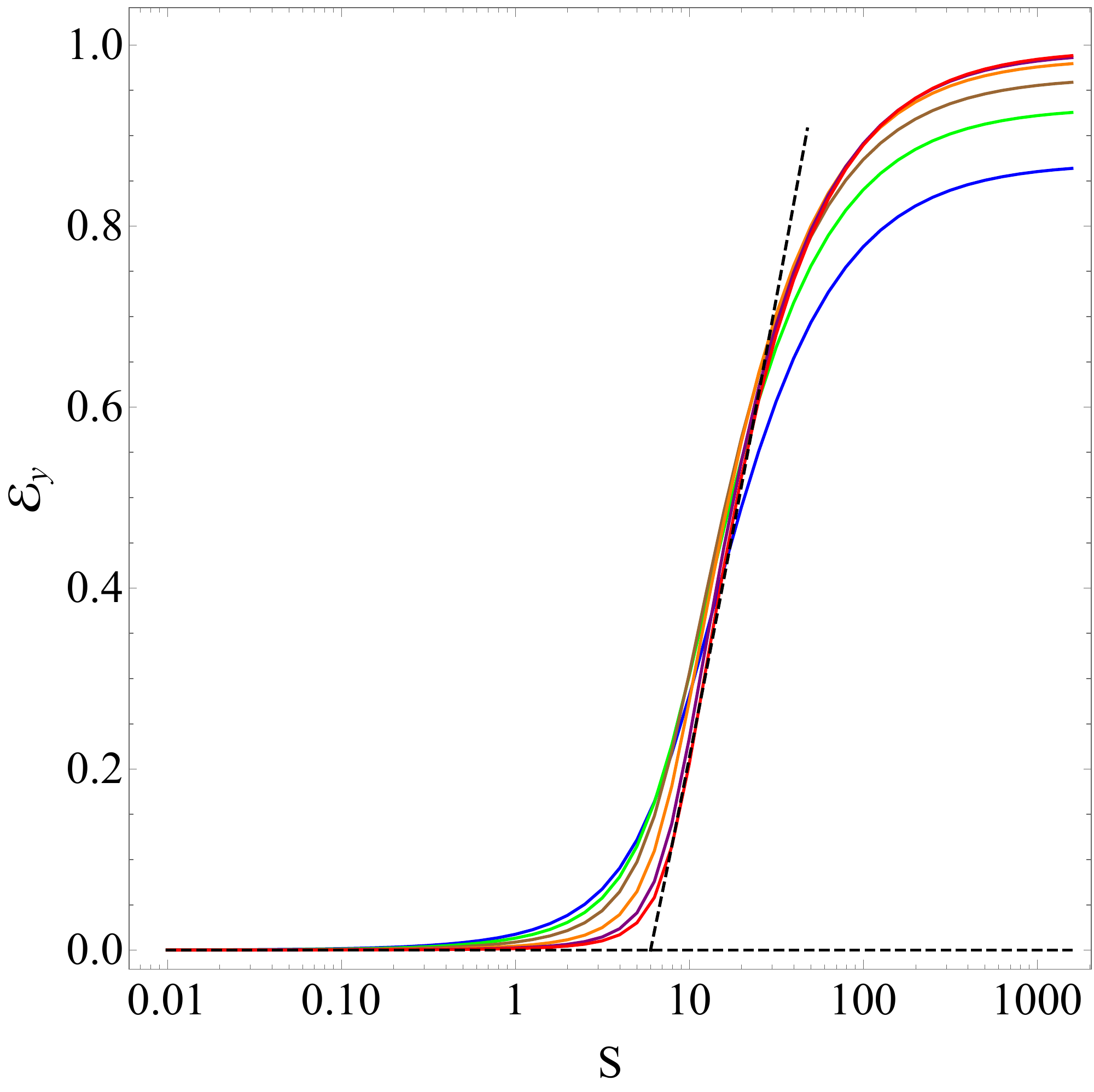} 
\hskip 1 truecm \includegraphics[scale=0.33,keepaspectratio=true]{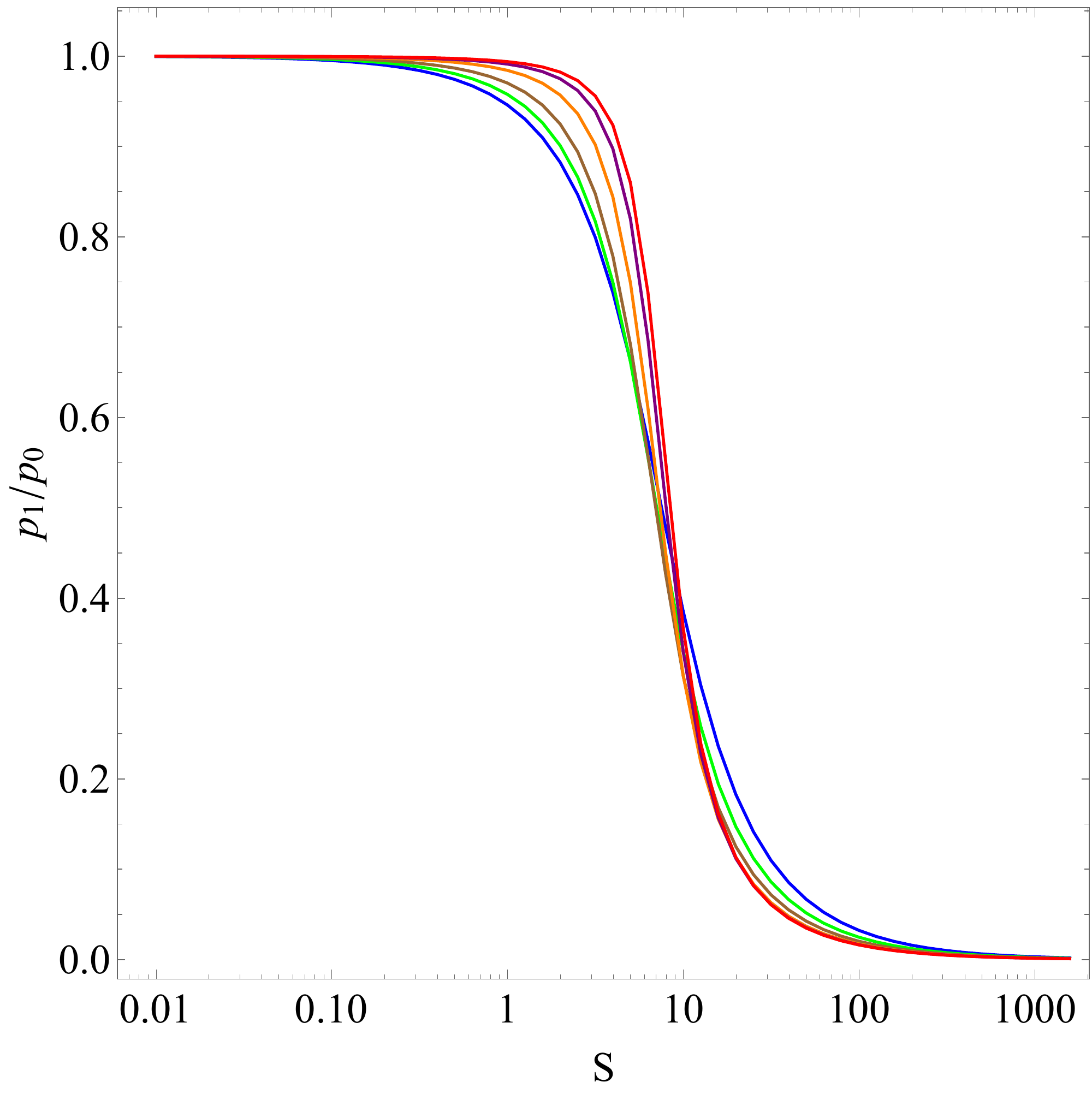} 
\vskip 0.35truecm
\includegraphics[scale=0.3,keepaspectratio=true]{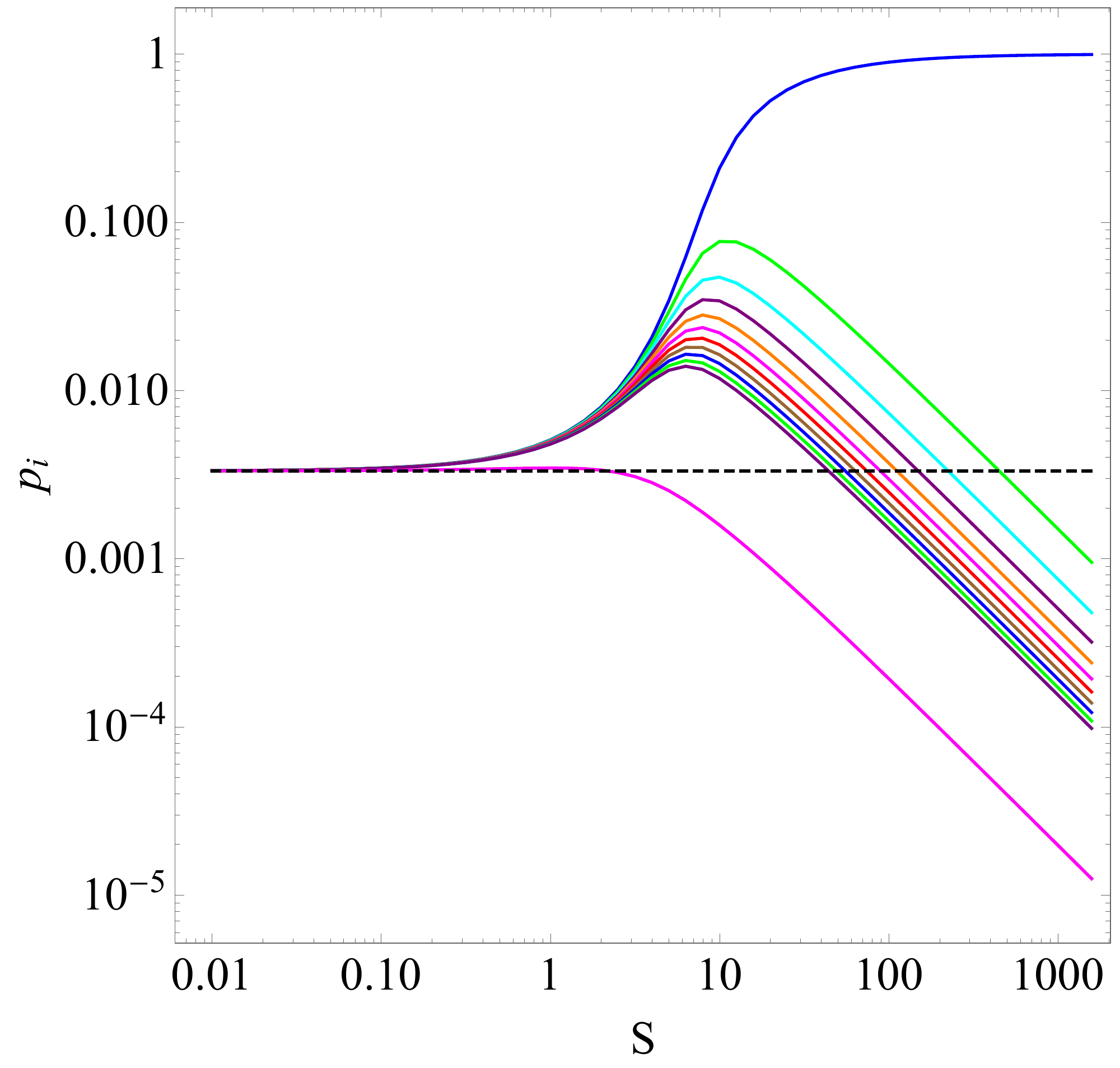} 
\caption{Classical Fr\"ohlich-like condensation. For an increasing number of modes entering the numerical simulations of Eqs.{\protect\eqref{eq:NonDim_ClassicalFrohlEqs}},  in the upper left panel the condensation index $\mathcal{E}_y$, defined in Eq.{ \protect\eqref{entropy}}, is reported versus the energy input rate $S$, and in the upper right panel the ratio of the energy contents $p_1/p_0$ is reported versus $S$. At equipartition $\mathcal{E}_0 =0$ and $p_1/p_0=1$; for the condensed state $\mathcal{E}_y = 1$ and  $p_1/p_0=0$.
The colours of the lines correspond to: $N_{sys}= 11$ (blue), $N_{sys}= 21$ (green), $N_{sys}= 41$ (brown), $N_{sys}= 101$ (orange), $N_{sys}= 201$ (purple), $N_{sys}= 301$ (red). The dashed oblique line (tangent to the inflection point of the two highest $N$ curves) is a guide to the eye to mark a possible asymptotic bifurcation point. 
In the lower panel the relative energy contents of two groups of modes: $p_i$ for $i = 0 - 10$ (from blue to purple) , and $i= 99, 100, 101$ (blue, green, magenta superposed) are displayed as a function of $S$, having set $N_{sys}= 301$. The dashed line corresponds to energy equipartition.}
\label{acondens}
\end{figure}

In Figs.\ref{fig:DiagrClasCondens_Sys2} and \ref{acondens} the results of numerical integration of the  rate equations \eqref{eq:NonDim_ClassicalFrohlEqs} are reported for the set of parameters
$B_{\omega_i}=B=1$, $C_{\omega_i \omega_j}=C=0.1$ with the initial conditions specified in the previous subsection. The coefficient $B$ has been chosen as reference parameter because it is directly related with the characteristic thermalization time scale $\tau_{\text{therm}}\approx B^{-1}$. The quartic coupling constant  has been chosen $\tilde{\Upsilon}_{(1)}=\tilde{\Upsilon}_{(2)}=10^{-4}$, which is sufficient to guarantee a finite bound from below of the energy. 
The choice $C/B=0.1$ is coherent with respect to what has been reported in the literature \cite{pokorny} concerning the already investigated qualitative aspects of quantum Fr\"ohlich condensation.
The integration has been performed using a fourth-order Runge-Kutta  algorithm, in each case for a time interval $\tau_{\text{int}}>0$ sufficiently long so as to guarantee that $|\epsilon_{\omega_i}(\tau)-\epsilon_{\omega_i}(\tau_{\text{stat}})|< 10^{-5}$  for $\tau_{\text{stat}}, t\in [0,t_{\text{int}}]$.
In Figure \ref{fig:DiagrClasCondens_Sys2} the reported results display the fraction of energy content in each mode for a system with $N=20$, that is $21$ normal modes and different values of the energy injection rate. It appears quite clearly that the larger the energy injection rate $S$ the stronger the deviation from energy equipartition with increasing energy concentration in the lowest frequency mode $\omega_{0}$. This energy concentration phenomenon, that we can call phonon condensation, stems from classical Liouville equation and displays several analogies with the phenomenology of the pseudo-Bose condensation predicted by Fr\"{o}hlich in a quantum framework, and subsequently largely studied and debated in the same quantum framework.

A relevant feature of the Fr\"{o}hlich condensation is the existence of a threshold in the energy injection rate which corresponds to the breaking of the above discussed permutation symmetry.  In order to find out if a similar threshold effect is displayed also by our classical model, we studied the behavior of the above defined indicators of symmetry breaking (the condensation index $\mathcal{E}_y$ and the ratio $y_{\omega_i}/y_{\omega_{0}}$) as a function of $S$ and for different values of $N$. Figure \ref{fig:DiagrClasCondens_Sys2} shows that,  for the chosen values of $S$, it is $p_{i}>p_{i+i}$ hence the parameter $p_{1}/p_{0}$ appears as a natural choice to detect a possible threshold effect. The results of numerical simulations of the order parameters have been reported in Figure \ref{acondens}. Both indicators pass from their respective equipartition values ($\mathcal{E}_y\simeq 0$, $p_1/p_0\simeq 1$)  to their condensation values ($\mathcal{E}_y\simeq 1$, $p_1/p_0\simeq 0$)  by displaying a clear steepening at increasing $N$. 
In particular, at increasing $N$, the curves of the condensation index $\mathcal{E}_y(S,N)$ display a tendency to superpose at large $S$-values, and a sharpening of the transition knee at lower $S$-values, suggesting an asymptotic convergence to a sharp bifurcation pattern. 
In fact, this sharpening of the transition curves with $N$ is reminiscent of the finite-size patterns of 
the order parameter as a function of the control parameter of equilibrium second order phase transitions. An analogous phenomenon is displayed by the $S$-pattern $(p_{1}/p_{0})(S)$ of the indicator $p_1/p_0$ at increasing $N$. Actually, the right panel of Figure \ref{acondens} shows that, around the value $S\simeq 10$, the different curves $(p_{1}/p_{0})(S,N)$ cross each other and get steeper at increasing $N$, again suggesting that the larger the numbers of modes the sharper the transition between almost-equipartition among the modes and phonon condensation in the lowest frequency mode.  It is worth noting that this phenomenology is qualitatively confirmed by the outcomes reported in Figure \ref{fgr:3} (aimed at  highlighting a saturation effect) where again a sharpening - at increasing $N$ - of a threshold effect can be observed. 
The lower panel of Figure \ref{acondens}, for the case of $N_{sys}= 201$, provides some interesting and more detailed information about the dynamics underlying the condensation phenomenon reported in the upper panels of the same Figure. In fact, the lower panel shows how the relative energy contents $p_i$ of a randomly chosen subset of modes ($p_0, p_1,  p_5, p_{10},  p_{100}$) deviate from energy equipartition at increasing $S$. We can observe that the highest frequency mode selected, $p_{100}$, deviates from equipartition at smaller values of $S$ than the lower frequency modes chosen; then, somewhat below $S=10$, the energy content of all the selected modes starts decreasing except the energy content $p_0$ of the lowest frequency mode which gathers more and more energy approaching the value $p_0=1$, that is, full condensation at high $S$.

The main result of this theoretical part of the paper is an original derivation of a classical counterpart of the well-known Wu-Austin quantum Hamiltonian from which Fr\"{o}hlich's rate equations can be derived. This classical counterpart consists of a system of nonlinear dynamical equations \eqref{eq:NonDim_ClassicalFrohlEqs} for the (adimensionalized) energy contents $y_{\omega_i}$ of each normal mode $\omega_i$ of an ensemble of oscillators coupled with two thermal baths at different temperatures. The ensemble of oscillators - ideally modelling the vibrations of the atoms or of groups of atoms in a macromolecule - is both linearly and nonlinearly coupled to the set of oscillators representing a thermal bath at temperature $T_{B}$; this thermal bath accounts for the environment (mainly water molecules) and for other fast internal oscillations of a macromolecule. The second thermal bath at a temperature $T_S\gg T_B$ models an external energy supply, in close analogy with the Wu-Austin model. 
The numerical study of Eqs. \eqref{eq:NonDim_ClassicalFrohlEqs} has shown that the condensation phenomenon is not peculiar of a quantum model, which a-priori was not an obvious fact, hence the occurrence of a phonon condensation phenomenon in a macromolecule at room temperature is more plausible on the basis of our classical model than on the basis of Fr\"{o}hlich's quantum model. 
Therefore, even though Equations  \eqref{eq:NonDim_ClassicalFrohlEqs} still represent a biomolecule in a very idealized way, the genericity of qualitative features of out-of-equilibrium phase transitions - mentioned at the beginning of this Section - suggests that the condensation phenomenon should actually occur in real biomolecules, even though our model is not expected to predict the quantitative details of the phenomenon. This is enough to make the predicted phenomenology worth an experimental effort to detect it. 

\section{Experimental detection of non-equilibrium collective mode}\label{manips}


Even though experimental evidence of the existence of collective modes of vibration of biomolecules has been provided  \textit{at thermal equilibrium} by means of Raman  spectroscopy \cite{painter} already many years ago, and is still being the object of many investigations \cite{turton,acbas,falconer,markelz,xu,ebbinghaus,lundholm}, no experimental evidence was hitherto available of the possibility of exciting out-of-thermal-equilibrium collective oscillations of a biomolecule. Unveiling whether these can be activated amounts to understanding whether  a necessary condition to activate long-range intermolecular electrodynamic forces \cite{Preto:2015}  can be fulfilled. This is what strongly motivated the joint theoretical and experimental work reported in the present paper. The experimental counterpart of the theoretical work relies on two complementary experiments.

\begin{figure}[h!] 
\centering
\includegraphics[scale=0.3,keepaspectratio=true]{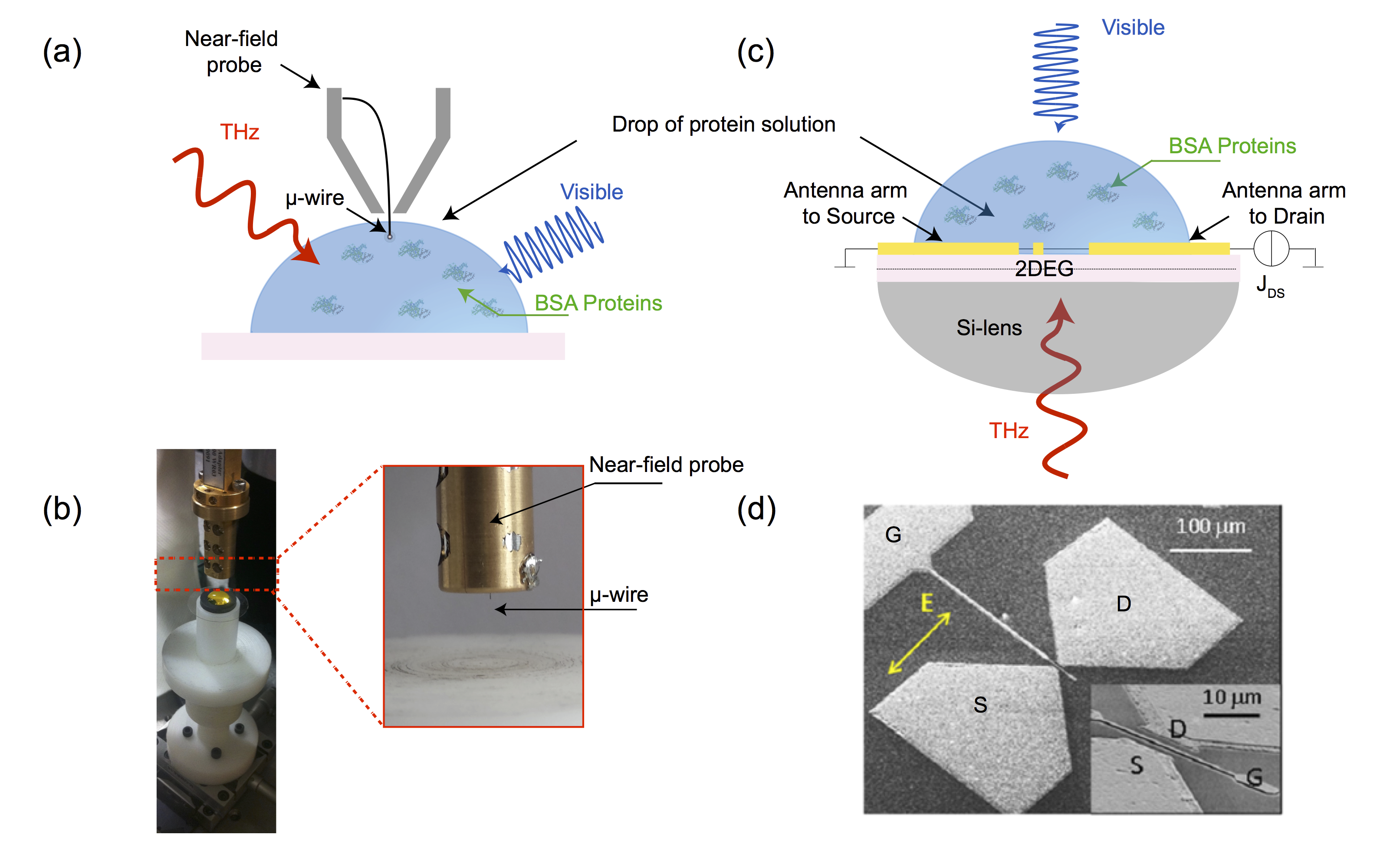} 
\caption{Experimental setups of THz absorption near-field spectroscopy. (a) A drop of the biological sample is placed under the near-field probe which is directly immersed inside the solution. (b) Picture of the near-field probe and its micro-wire. (c) A drop of the biological sample is placed above the near-field rectenna. (d) Electron-beam microscopy picture of the bow-tie antenna with its integrated FET.}
\label{fgr:1}
\end{figure}

For both experiments a model protein has been chosen: the BSA (Bovine Serum Albumine) protein. This is mainly made out of $\alpha$-helices, and is a  "model" since it is largely studied in the biophysical chemistry  literature. Our strategy to create 
a stationary out-of-thermal-equilibrium state of this molecule is to induce it by means of optical pumping,  without involving any optical transition of the protein,  through the excitation of  some fluorochromes bound to each protein molecule. The optical excitation of these fluorochromes creates on each protein some ``hot points''  acting as the epicentres of a so-called ``proteinquake'' \cite{ansari,levantino}  - better discussed in the following - and resulting in an energy transfer to the vibrational part of the protein. We used the Alexa488 fluorochrome which is covalently bonded at the lysine residues of the BSA and which is excited by means of an Argon laser (wavelength $488$ nm). Some $0.19$ eV per fluorochrome and per incident photon (the average energy difference between the absorbed and re-emitted photons) is thus available for an energy transfer to the protein and, partly, also to its environment.
 By attaching an average number of 5 fluorochromes per protein a considerable amount of energy ( $\gg k_BT$) can be continuously pumped into each protein. 
Two THz-near-field absorption spectroscopy setups of aqueous solutions of the protein (at 1 mg/mL concentration) operating into two distinct laboratories, have been used at room temperature (Fig. \ref{fgr:1} (a), (c)). In both experiments THz radiation is produced by tunable, highly-spectrally-resolved ($< 300$ Hz) and continuous-wave sources with an average power of 1~mW allowing an accurate detection of possible resonances. A typical experiment consists in three phases, during all of them the aqueous solution of proteins is illuminated with THz radiation performing a sweep in frequency, and thus allowing to measure the frequency dependence of the absorbed electromagnetic power (detected by the near-field probes) in the solution. During the first phase no extra illumination with the Argon laser is done; during the second phase the Argon laser is switched on to excite the fluorochromes bound to the proteins; finally, during the third phase the laser is switched off to check whether some memory and irreversible photochemistry effect (photobleaching)
or sample heating are  present. The use of near-field coupling of metal probes to the sample eliminates the Fabry-Perot interferences often seen in optical spectra taken in fluidic cells \cite{globus}.

The first setup (Fig. \ref{fgr:1} (a), (b)) used a micro-coaxial near-field probe put inside a metallic rectangular waveguide enabling a modal transition from $TM_{01}$ Sommerfeld's  to $TE_{01}$ waveguide mode. The sub-wavelength diameter of the wire (12 $\mu$m) allows an extremely localized detection of the longitudinal component of the electric field at its end and on a volume of about 4 pL.   The spectra were subtracted from the spectrum of pure water in order to remove artifacts coming either from the water absorption or from the geometry of the experimental setup. 
The second setup \cite{giliberti}  (Fig. \ref{fgr:1} (c), (d)) used a near-field probe rectenna composed of a planar metal bow-tie antenna with dimensions close to half-a-wavelength (at 0.3 THz) that enhanced the THz field in the feed gap region (volume of about 0.2 pL) and a plasma-wave field-effect transistor (FET) integrated in the feed gap of the antenna. When illuminated by THz radiation, the antenna-coupled FET device provides
a DC-voltage -- between Source and Drain contacts -- proportional to the THz-field intensity  \cite{dyakonov,knap,nouvel}. A hemispherical silicon lens pressed on the back of the semiconductor substrate focused the THz radiation on the antenna simultaneously eliminating Fabry-Perot interference in the substrate \cite{digaspare}.  The spectrum of the protein solution obtained in the absence of blue light  illumination was subtracted from the spectrum obtained with blue light illumination.

\begin{figure}
\centering
\includegraphics[scale=0.65,keepaspectratio=true]{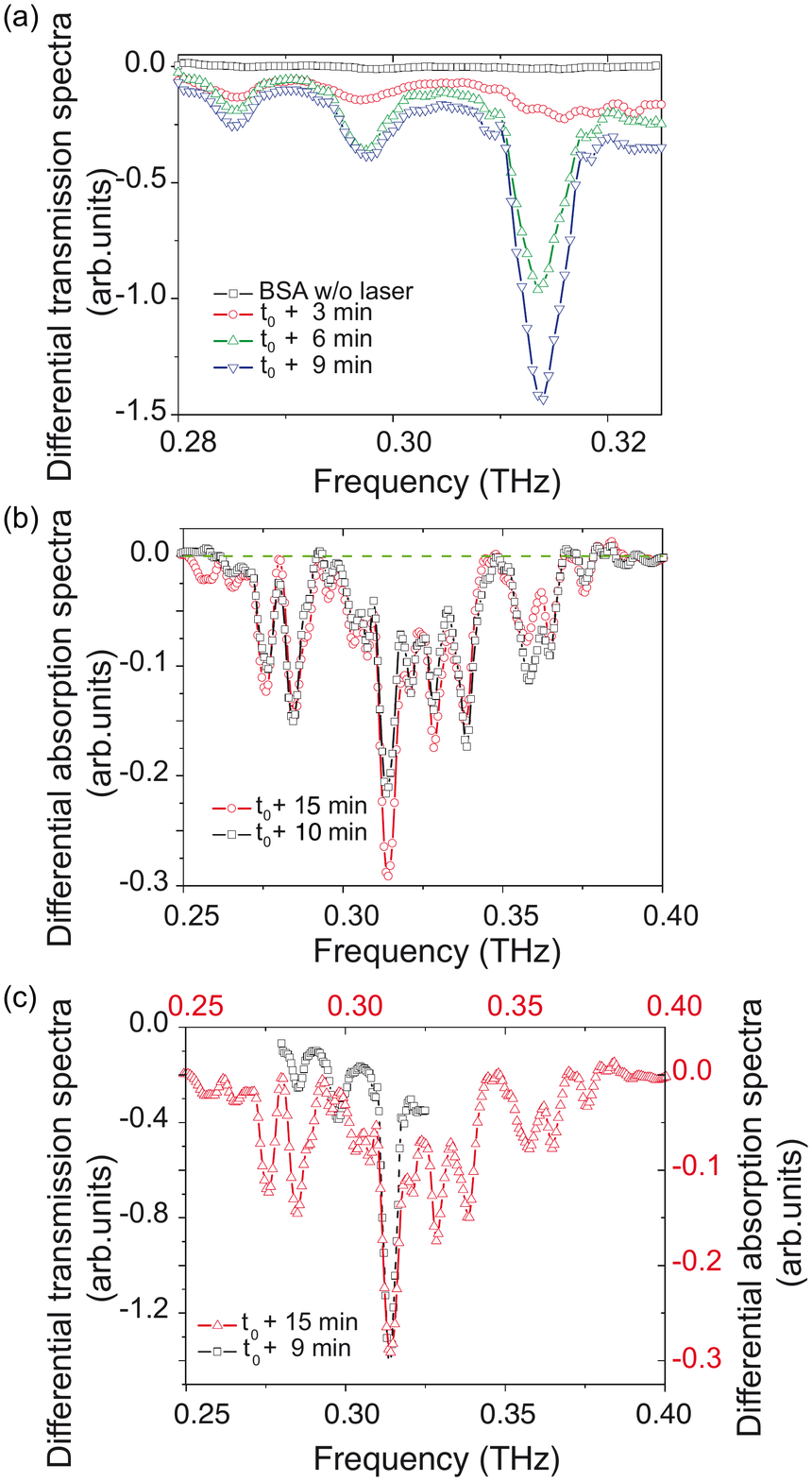} 
\caption{Differential transmission and absorption spectra as functions of the frequency. (a) Spectra obtained using the microwire probe after subtraction of the water spectrum in the absence of illumination and in the presence of illuminations for the reported durations. (b) Spectra obtained using the rectenna, after subtraction of the protein solution without illumination, for the reported durations. (c) Comparison of the two normalized spectra for the longest illumination durations.} 
\label{fgr:2}
\end{figure}

Figure \ref{fgr:2} (a) presents the spectra obtained using the micro-coaxial probe in the absence of blue light  illumination (black squares) and in the presence of blue light  illuminations of different durations (from 3 to 9 min). In the former case (no illumination), there was no specific spectral feature in the studied frequency range while in the latter case (with illumination) we observed spectral resonances which become more evident for increasing duration of illumination. In particular, the strongest resonance appeared at 0.314 THz accompanied by two other minor resonances  situated at 0.278 and 0.285  THz; these values did not depend on the time of illumination and the strength of the resonances saturated after 9 min of illumination.  These  results have full reproducibility.
Figure \ref{fgr:2} (b) presents the spectra obtained using the rectenna probe for two durations of blue light  illumination. The spectra with illumination-on versus illumination-off were taken several times (two of them reported to show reproducibility). Also in this case we observed the appearance of evident resonances whose strength saturated at increasing durations of illumination (10 and 15 minutes in the two experimental runs reported). 
Long excitation times are needed because, under our experimental conditions,  the energy dissipation rate and the energy supply rate are almost equal, so that a long time is needed to accumulate enough energy into each protein in order to make intramolecular nonlinear interaction terms sufficiently strong to activate the condensation phenomenon.

The spectra obtained using the two previously described methods and for the longer durations of illumination are compared in Fig. \ref{fgr:2} (c). The principal resonance at 0.314 THz is perfectly reproduced using two completely different and complementary setups. Since THz extinction in water is huge ($2000$ dB/cm), the emergence of this spectral feature of the protein out of the water absorption background must be associated with the activation of a giant dipole moment. And this can happen only as a consequence of the activation of a coherent oscillation of the whole molecule, possibly together with a dipole moment strengthening  due to an electret-like structuring of water dipoles of the first hydration layers of the proteins as hydration shell might contribute to the  magnitude of the protein dipole moment \cite{susko}. 

Since the BSA is a heart-shaped  globular protein, a possible lowest frequency {of a global oscillatory mode is roughly estimated by schematising} the molecule as composed of two masses $m$, equal to half the total protein mass, joined by a spring of elastic constant $k$ given by $k=E A_0/l_0$, where $E$ is the Young modulus of the protein, $A_0$ and $l_0$ are its transverse section and length at rest, respectively. Using $m=33$ kD, $A_0\simeq 1.2\times 10^{-13}$ cm$^2$, $l_0\simeq 1.2\times 10^{-7}$ cm, and $E= 6.75 \ \rm{GPa}$, we find $\nu = (1/2\pi) \sqrt{k/m}\simeq 0.300$ THz which is close to the main resonance observed at $0.314$ THz. Since the BSA molecule can be modelled to first order as a three-dimensional elastic nanoparticle \cite{xu}, a more refined approximation is obtained by modelling the protein with an elastic sphere and then considering its vibrational frequencies. 

The fundamental frequency of a spheroidal deformation mode of an elastic sphere is given by the formula \cite{bastrukov}
\begin{equation}
\nu_0 = (1/2\pi)[2(2l+1)(l-1)]^{1/2}\left(\frac{E}{\rho R_H^2}\right)^{1/2}
\label{collVib}
\end{equation}
which holds for $ l\ge 2$. Using the following data for the BSA protein: Young modulus $E= 6.75 \ \rm{GPa}$ obtained at room temperature using Brillouin light scattering of hydrated BSA proteins \cite{perticaroli}, hydrodynamic (Stokes) radius $R_H=35 \textrm{\AA}$, and specific volume $1/\rho = 0.74$ derived from small-angle X-ray scattering (SAXS) experiments 
\cite{mylonas}, we find for the lowest mode ($l=2$) the frequency $ \nu_0 = 0.308 \ THz$ which agrees within an error of about $2\%$ with the observed peak value at $ \nu = 0.314 \ THz$. Though such a modeling is unrealistic in what it does not take into account the details of the protein structure and the associated normal modes \cite{suhre}, it nonetheless catches a  relevant aspect of the global deformation dynamics of the BSA molecule, namely the activation of a collective oscillation, also suggesting that the physical parameters adopted correspond quite well to the situation investigated {\cite{bosonpeak}. }
Secondary resonances are also present in both spectra. A possible explanation could be tentatively given considering torsional modes. These could be activated at the frequencies given by the relation \cite{bastrukov}
$$
\nu_t = \nu_0 \left(\frac{(2l+3)}{2(2l+1)}\right)^{1/2}\ , \quad\quad\quad\quad l\ge 2
$$
where $\nu_0$ is given by equation (\ref{collVib}), whence, for $l=2$ and $l=3$, one finds $ \nu = 0.257 \ THz$ and $ \nu = 0.246 \ THz$ respectively. These could be associated with  the two weaker absorption lines observed at $ \nu = 0.278 \ THz$ and $ \nu = 0.285 \ THz$. Here the larger discrepancy can be attributed to the non-spherical shape of the BSA, what entails the existence of different  moments of inertia according to the rotation axis, whereas the breathing mode is insensitive to this fact. Minor peaks are observed at higher frequencies with the rectenna (falling outside the accessible frequency range of the near-field probe) when the protein solution is illuminated with blue laser light. However, the blue light illumination could produce spurious signals from the 2D electron gas of the FET junction as a consequence of electron-hole pairs excitation causing a change of the transistor channel conductivity. This effect is well known and studied in the literature \cite{bubanga} so that minor peaks could be instrumental artifacts due to this electron-hole pair creation effect.

Let us stress an important point: computational normal mode analysis for proteins has shown nearly continuous vibrational density of states \cite{suhre} which have also been proved nearly uniformly optically active. Moreover, the coupling of these vibrational modes with water results in broad absorption features \cite{heyden,sun}. But this is true at \textit{thermal equilibrium}, whereas under out-of-equilibrium phonon condensation the energy content of all the normal modes is strongly depleted with the exception of the collective mode. Whence a narrow absorption feature. And, in fact, the computation of the function $L(\omega)$ in Section \ref{materials} shows how a dipole actively oscillating at a given frequency entails an absorption feature of shape similar to the experimentally observed one.  

\begin{figure}[h!] 
\centering
\includegraphics[scale=0.3,keepaspectratio=true]{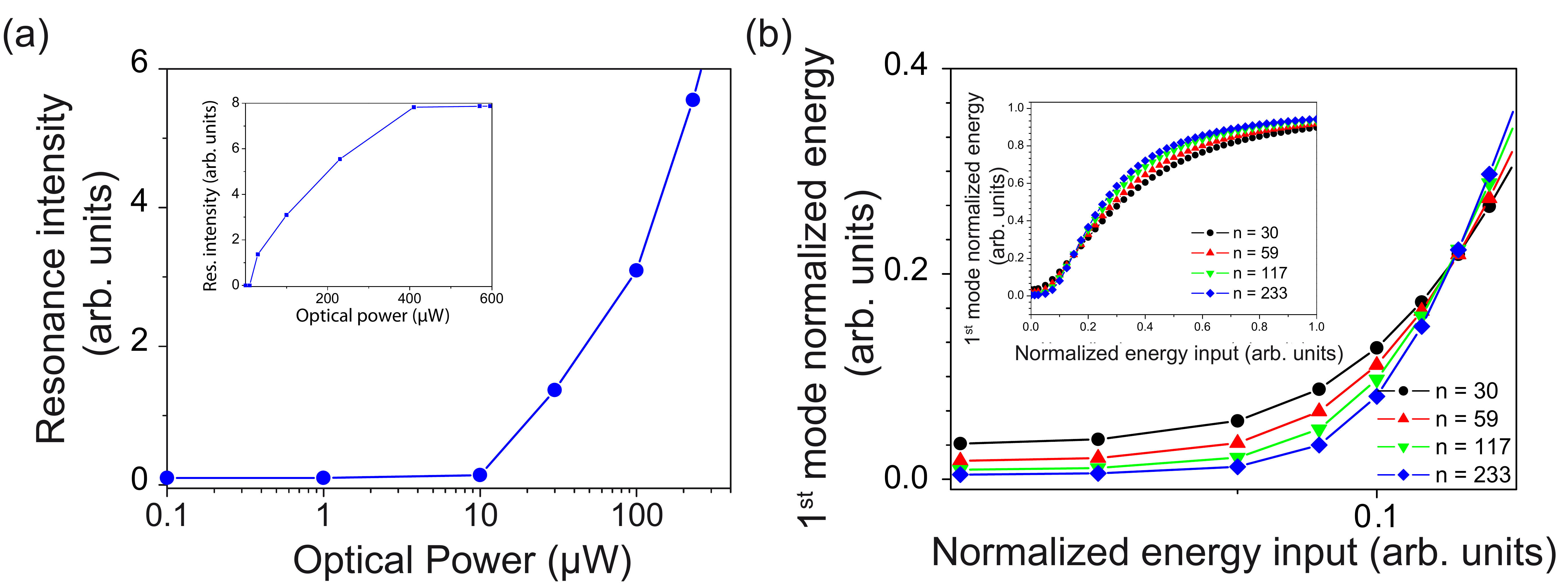} 
\caption{Threshold-like behaviour of giant dipolar oscillations. (a) Intensity of the resonant peak measured at $0.314$ THz as a function of the optical laser power. (b) Normalized energy of the fundamental mode calculated as a function of the normalized source power. The different curves correspond to the reported numbers of normal modes of the BSA protein. Theory and experiment are in \textit{qualitative} agreement.}
\label{fgr:3}
\end{figure}

According to our classical version of the Fr\"ohlich model, it is also expected that the appearance of a collective oscillation should exhibit a threshold-like behaviour when increasing the energy flowing through the protein. Actually, Figure \ref{fgr:3} (a) presents a clear threshold in the intensity of the resonance peak at $0.314$ THz when the optical input power exceeds 10 $\mu$W. By using a classical formalism for the analysis of the out-of-equilibrium phonon condensation we have calculated the intensity of the normal vibrational  modes of the BSA-protein as a function of the source power injected through the protein. Figure \ref{fgr:3} (b) highlights a threshold-like behaviour of the intensity of the fundamental mode that accumulates the energy at the expenses of the excited modes, in \textit{qualitative} agreement with the experimental outcome. By increasing the number of modes included in the calculation this threshold becomes more and more evident.  The experimental and theoretical results reported in Figure \ref{fgr:3} agree also in displaying a saturation effect occurring at large values of the energy input rate.

The observed spectra are certainly due to the light-excited protein because the spectral feature at $0.314$ THz was not observed by illuminating: {\it i)} water alone; {\it ii)} a aqueous solution of the fluorochrome in the absence of the protein; {\it iii)} a aqueous solution of the BSA protein without the bound fluorochrome (see Section \ref{materials}). On the other hand, the observed spectral line at $0.314$ THz immediately disappears by switching off the laser. Remarkably, the spectra obtained with two independent and different experimental setups, based on two different methods of detection  of the THz radiation, operated in two different laboratories, show a strikingly good overlap of the respective absorption line profiles at $0.314$ THz. A result in excellent agreement with the frequency of $0.308$ THz predicted for the spheroidal (collective) vibrational mode computed on the basis of recent experimental measurements of the relevant parameters of the BSA protein. This triple concordance among the results so far obtained should be enough to exclude experimental artifacts. This notwithstanding various sources of artifact for the observed phenomenology were considered.  
A first objection suggests itself, namely that  the observed phenomenology is just a trivial heating effect due to the laser light. This would be true in the absence of a frequency dependent response to the injection of energy into the proteins. Heating indeed means increasing of the kinetic energy of the atoms and group of atoms of the protein entailing energy equipartition among the vibrational modes of the protein. Hence neither a frequency dependent effect nor a threshold effect for the energy input rate would have been measured, but just a new thermal equilibrium state would have been achieved. To the contrary, each protein - submitted to continuous energy feeding and energy dissipation - behaves as an open system undergoing a non-equilibrium phase transition: when the ratio between energy gain and losses exceeds a critical value a collective behavior sets in producing the phonon condensation, as it has been already discussed in Section \ref{classicBose}. Then another question arises about the conversion mechanism of the visible light energy absorbed by the electrons of the complex protein/dyes into the vibrational modes of the proteins. This mechanism of rapid intramolecular dissipation of energy through quake-like structural motions as a consequence of a perturbation (such as the breaking of a chemical bond or the absorption of photons through electronic transitions) is being given increasing experimental attention \cite{ansari,levantino,arnlund,brinkmann}  and is referred to as ``protein quake''. Similarly to an earthquake, this effect describes how a protein strain is released at a focus or ``hot-point'' (in our case the fluorochromes) and then rapidly spreads as a structural deformation through waves, thus exciting protein vibrational modes. Another source of artifact could be the apparition of standing waves and related interferences, but  these would have been easily identified. Furthermore, there would be no reason for such interferences - if any -  to manifest themselves as a consequence of the excitation of fluorochromes through the blue light [see Figure (\ref{laser-off}b)].

\section{Materials and methods}\label{materials}

\subsection{Sample preparation}
Bovine Serum Albumine (BSA) was purchased from Sigma-Aldrich (A7030) (St. Louis, MO) in lyophilised powder, protease free, fatty acid free ( $\le 0.02\%$), essentially globulin free; being the degree of purity of the BSA used higher than $98\%$, it was used without further purification and was dissolved directly into bidistilled water. 
The BSA molecules have been labeled with the fluorochrome AF488 5-TFP (Alexa Fluor $488$ Carboxylic Acid, $2,3,5,6-$Tetrafluorophenyl Ester), $5$-isomer (A$30005$) which was purchased from Molecular Probes Invitrogen.

The dye has excitation/emission of 495/515 nm and a molar extinction coefficient of $ \varepsilon_{495}  = 71000$ M cm$^{-1}$. Both the protein and dye concentrations have been determined measuring their absorbance with a Nanodrop 1000 Spectrophotometer (ThermoScientific), at $280$ nm with a molar extinction coefficient of $36000$ M cm$^{-1}$ and at 495 nm with a molar extinction coefficient of $71000$ M cm$^{-1}$, respectively.  The chemically labeled BSA molecules have been obtained by homemade labelling using 2 mg/mL aqueous solution concentration of proteins with an initial ratio of concentrations [A488 5-TFP]/[BSA] equal to 8  in sodium bicarbonate at a pH $\approx$ 8.5, during one hour. Unconjugated dye was removed using a PD-mini-trap G25 (GE Health Care) according to the instructions of the manufacturer, using gravity protocol, and the degree of labeling was determined spectroscopically. After purification each BSA molecule has between 5 and 6 fluorochromes attached.

\subsection{Experimental detection of the collective mode}
Two separated and different experiments, performed in Montpellier (France) and in Rome (Italy), respectively,  have been performed to get the terahertz non-equilibrium spectra of the model protein chosen. The former used a microwire as local probe whereas the latter used a nanorectenna.

\subsubsection{Microwire-based THz spectroscopy}

A constitutive element of the experimental setup is a tunable and continuous-wave primary source from Virginia Diodes Inc.(https://www.vadiodes.com/en/) thus emitting in the 0.22-0.33~THz frequency range with an average power of 1~mW. 
The high spectral resolution ($< 300$ Hz) of the continuous-wave source allows an accurate detection of resonances. The emitted radiation beam is focused on the samples of protein solutions on which, at option, a 488 nm light beam -- produced by an Argon laser -- can also be focused. The blue light provides the proteins with the necessary energy to activate a collective vibrational state. The latter being an out-of-equilibrium state because it is kept by a non-thermal energy supply.

The THz near-field scanning spectroscopy technique in aqueous medium is performed by resorting to a homemade micro-coaxial (i.e. subwavelength) near-field probe put inside a metallic rectangular waveguide connected to an heterodyne head and an electrical spectrum analyser. The subwavelength diameter of the wire allows an extremely focused enhancement of the longitudinal component of the electric field at its end over a volume of 4 pL.
In 1995, F. Keilmann \cite{keilmann} highlighted the advantages of introducing a metal wire in a circular metal waveguide to produce probes for near-field microscopy in the far--infrared and microwaves frequency domains. The advantage of this method is to avoid the frequency cut--off when the diameter of the guide is a subwavelength one. The circular waveguide is transformed into a coaxial waveguide which does not have low frequency cut-off  and which makes superfocalisation and high--resolution imaging possible.

However, for the experiments reported in the present work, a rectangular waveguide was used, and this entails some different phenomena since the micro-wire is soldered along the long axis of the waveguide and bent to exit it. This has two main consequences: 
on the one hand, the micro-wire enables a modal transition and, on the other hand, it serves as a waveguide. The bent portion of the micro-wire allows the conversion of the fundamental mode $TE_{01}$ inside the rectangular waveguide into a $TM_{01}$ mode along the wire. 
To optimize the coupling efficiency between the near-field and the probe, a special care has been first paid to the positioning of the micro-wire inside the guide. More precisely, the maximum of the near field signal is theoretically attained for a wire positioned at $L_{inside} = p \frac{\lambda_{TE_{01}}}{4} $ from the open side of the waveguide, where $\lambda_{TE_{01}}$ is the wavelength of the $TE_{01}$ mode at the considered frequency. At 0.3~THz, the micro-wire must be fixed at an entire multiple $p$ of 250~$\mu$m. The best compromise between efficiency and technical possibilities was found for $p=4$ that gives $L_{inside}=1$~mm.
The second parameter to take into account is the angular positioning of the bent portion of the micro-wire inside the rectangular waveguide. The coupling is maximum when the wire is parallel to the orientation of the electric field in the guide, that is, along the long-axis of the rectangular waveguide. Finally we also paid attention to two relevant parameters that are the total length and the diameter of the wire. The intensity of the electric field is roughly sinusoidal and its maximum is attained when the total length $L$ is a multiple of the half-wavelength. The best compromise between theory and technological possibilities has given a total length of $L= 2$~mm. Since the electric field intensity at the micro-wire extremity increases when the diameter decreases we used a wire of 12~$\mu$m diameter. All the previously mentioned parameters have also been simulated using CSTmicrowaves studio$^{\textregistered}$
(https://www.cst.com/products/cstmws) to ensure the better coupling efficiency as possible.
 
\begin{figure}[h!]
\centering
\includegraphics[scale=0.50,keepaspectratio=true]{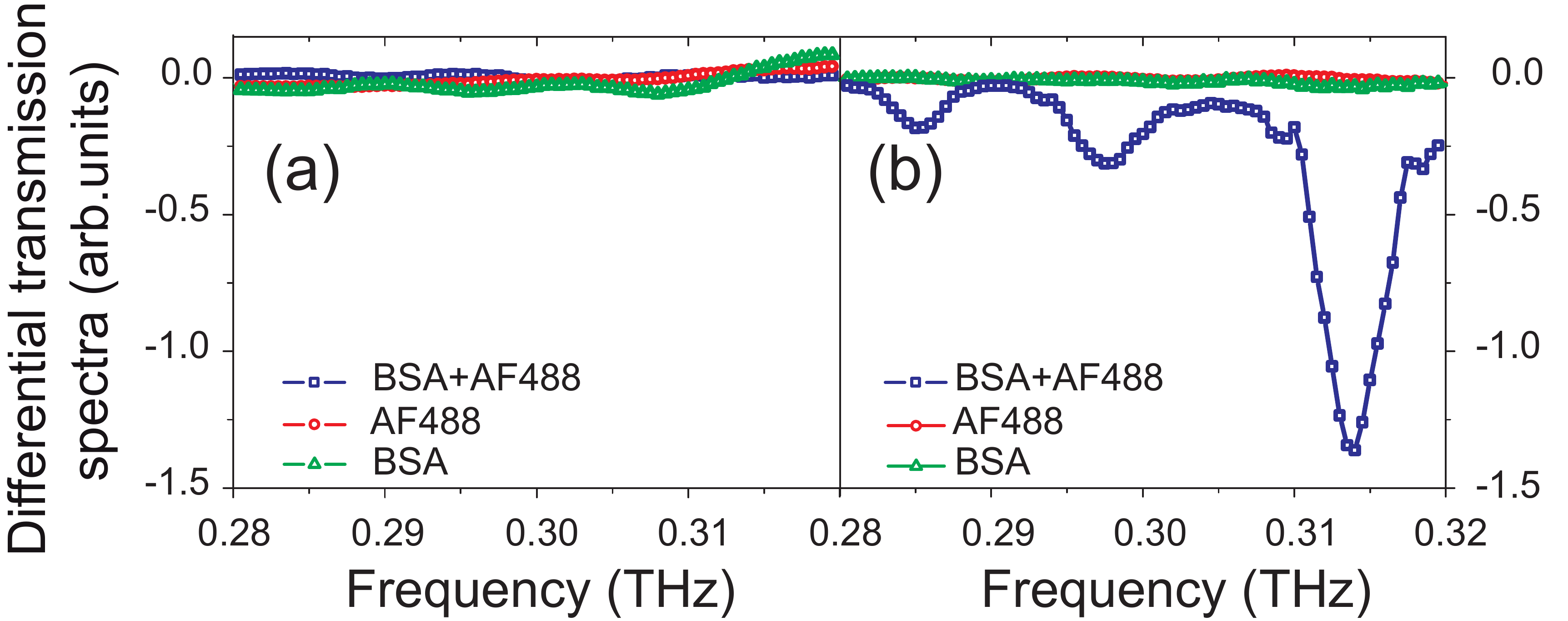} 
\caption{Differential transmission spectra of the microwire-based absorption spectra. (a) Without blue light illumination (laser OFF). Obtained with: solvated AF488 in water (red circles); solvated non-labelled BSA in water (green triangles); BSA labelled with AF488 solvated in water (blue squares). (b) With blue light illumination (laser ON). Obtained with:  solvated AF488 in water, at a concentration of 5 mg/mL  (red circles); solvated non-labelled BSA in water (green triangles); BSA labelled with AF488 solvated in water (blue squares). Labelled and non-labelled BSA concentrations were kept equal to 1 mg/mL.}
\label{laser-off}
\end{figure}
\begin{figure}[h!]
\centering
\includegraphics[scale=0.4,keepaspectratio=true]{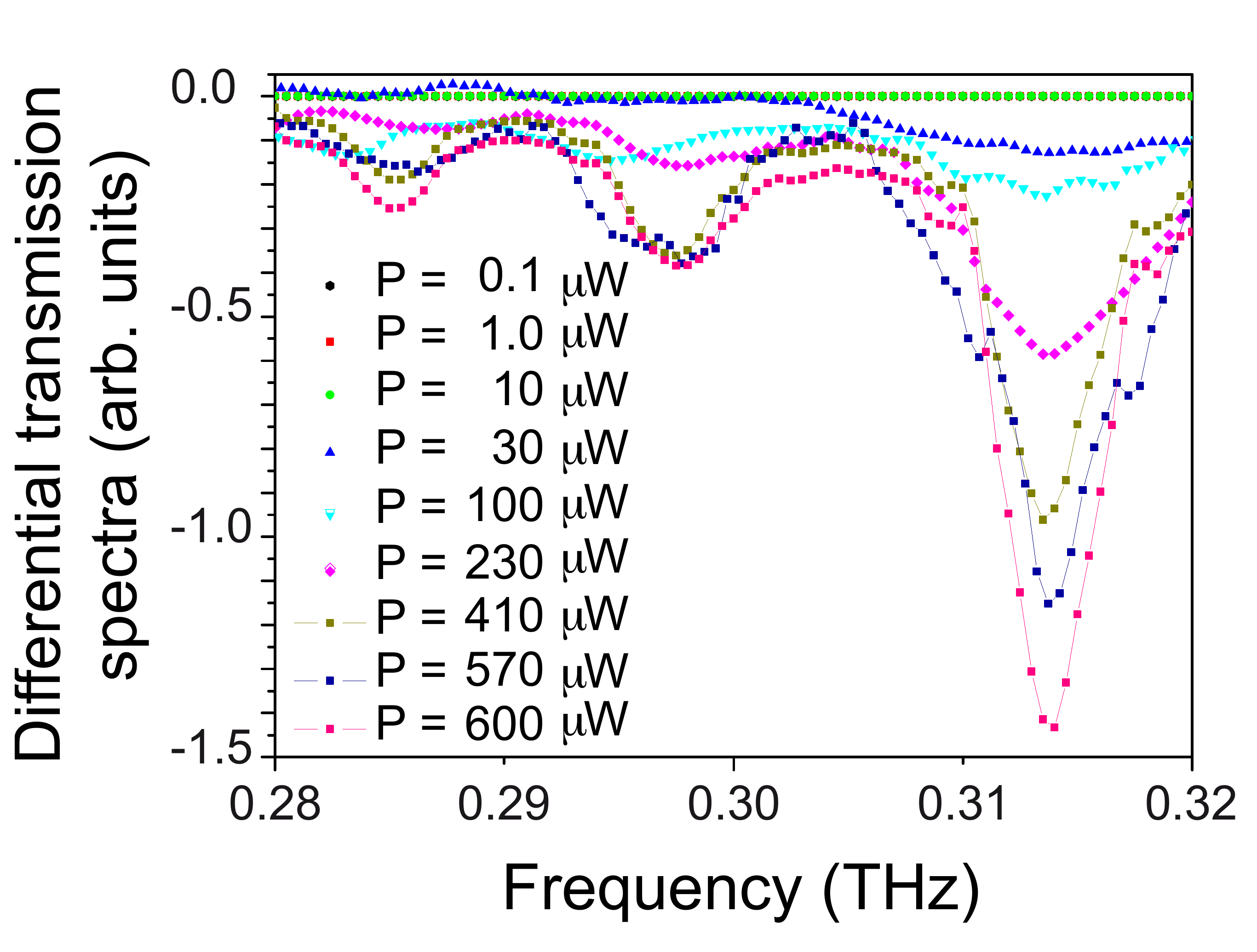} 
\caption{Differential transmission spectra of the microwire-based absorption spectra obtained with different excitation powers of the laser. The labelled BSA concentration was kept fixed at 1 mg/mL. The first three spectra are exactly superposed.} 
\label{fgr:7new}
\end{figure}

All the measurements have been performed at room temperature. A drop of the protein solution sample is placed under the near-field probe which is directly immersed inside the solution. A typical experiment consists in measuring the near-field electric field intensity through a reference medium (water) and the protein solution sample. A sweep through the frequency domain accessible to the THz source is performed alternatively when the blue light is switched-on and off to yield a difference spectrum showing the absorption features that are attributed to collective vibrations of each protein molecule.  

{Figure \ref{fgr:7new} displays the spectra, stabilised in time, obtained by varying the power of the laser illuminating the solution of labelled proteins at the constant concentration of  1 mg/mL. Notice that the secondary peaks appear and develop together with the main peak at 0.314 THz therefore also these secondary peaks are vibrational  modes of the entire protein, giving some support to the hypothesis attributing them to torsion modes. However, these kind of modes cannot be accounted for by our theoretical model because it contains no explicit information about the geometry of the molecule.}

\subsubsection{Rectenna-based THz spectroscopy}
In order to minimize the optical depth of water, in the second experiment (Rome) the probe domain was reduced to a volume of $10\times 10$ microns in $xy$ (horizontal plane), and to about $2$ microns in $z$ (vertical axis). To confine the THz radiation  (wavelength $\lambda$ around 1 mm) to such a deeply sub-wavelength region, a plasmonic antenna is used \cite{giliberti}. This device is based on two main components:  a planar metal antenna with length close to $\lambda /2$ (bow-tie, broadband type) that produces a high THz field region with antenna feed gap of $10\times 10$ microns; and a plasma wave FET transistor, which is a THz nonlinear electronic device integrated in the feed gap of the antenna, that provides an electric signal proportional to the THz field strength in the feed gap only. The plasma wave transistor was first introduced by Dyakonov and Shur in 1993  \cite{dyakonov} and further developed by many authors \cite{knap}, the main advantage being that standard microwave transistor technology can be employed for fabricating a device which is sensitive to THz radiation. The device is mounted in a package with a silicon lens pressed on the back of the semiconductor substrate and illuminated from below with a tunable THz oscillator (by Virginia Diodes Inc., 0.18 to 0.4 THz) through a set of off-axis parabolic mirrors. The resolution of the free-running oscillator is 2 GHz. After acquiring the empty-channel response spectrum of the device, a drop of protein solution was cast on the top (air) side of the device with a micropipette (a volume of 1 microliter was drop-casted). The drop extends over the entire antenna, but the radiation comes from below (i.e. from inside the substrate) and it is not attenuated by the whole drop.
In this experiment absorption from molecules outside the antenna feedgap in the $xy$ directions can be disregarded. Also, in the $z$ direction the field extends for less than 2 microns due to the plasma wave properties (the extension of plasmonic field in the $z$ direction is calculated by Finite Element Modeling simulations). Therefore, the number of molecules probed in this setup is that present in a $10\times 10\times 2$ micron volume, $10^{-6}$ times less than the number of molecules present in a $1\ mm^3$ diffraction limited focus. 
\begin{figure}[h!]
\centering
\includegraphics[scale=0.35,keepaspectratio=true]{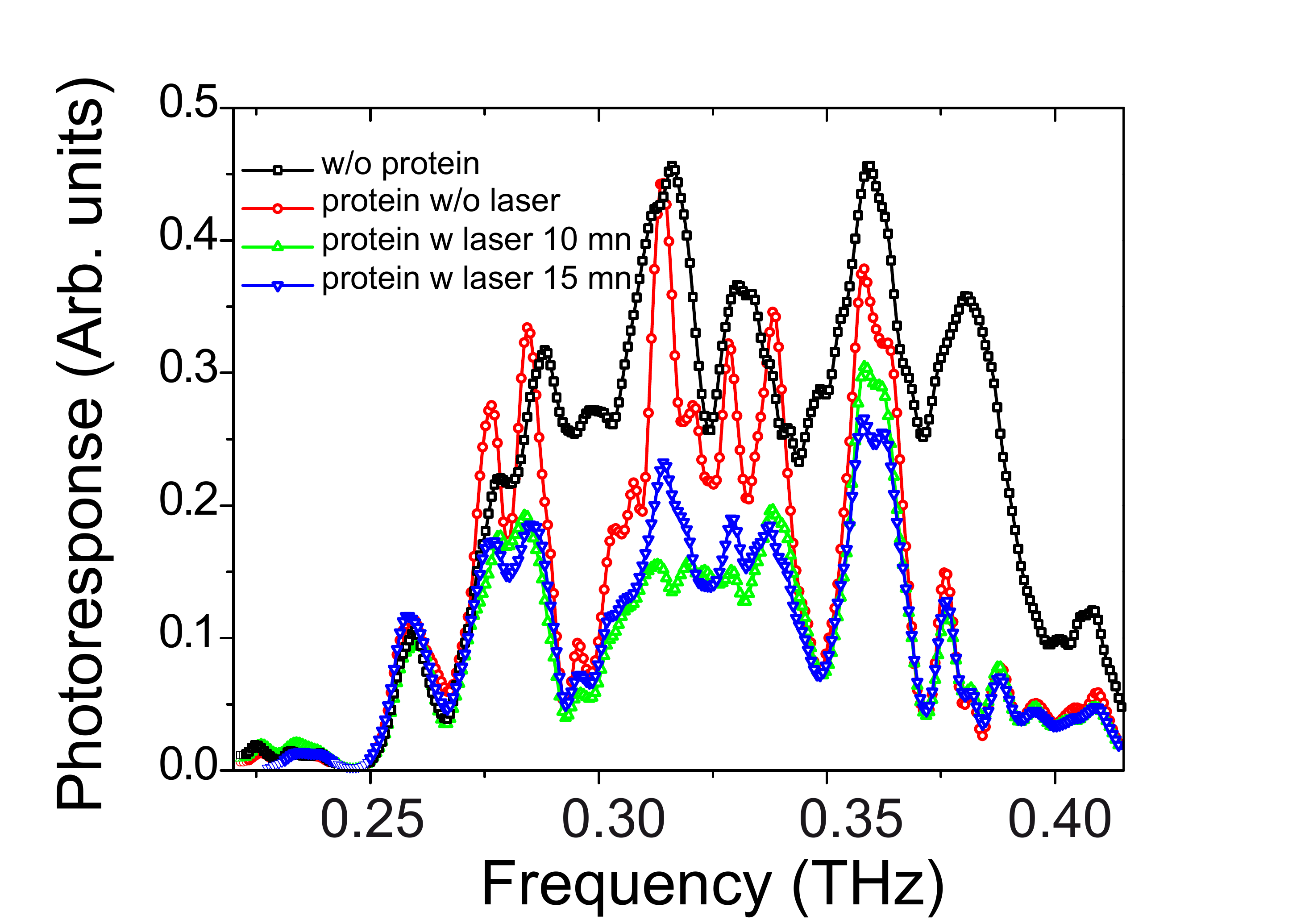} 
\caption{Raw data of the rectenna-based absorption spectra. Black squares correspond to the THz emitting source spectrum without the protein solution. Red circles correspond to the absorption spectrum of the BSA protein solution without blue light illumination. Green triangles correspond to the absorption spectrum of the BSA protein solution recorded after 10 minutes of blue light illumination. Blue triangles correspond to the absorption spectrum of the BSA protein solution recorded after 15 minutes of blue light illumination.
BSA concentration equal to 1 mg/mL.}
\label{romespectrum}
\end{figure}
\begin{figure}[h!]
\centering
\includegraphics[scale=0.35,keepaspectratio=true]{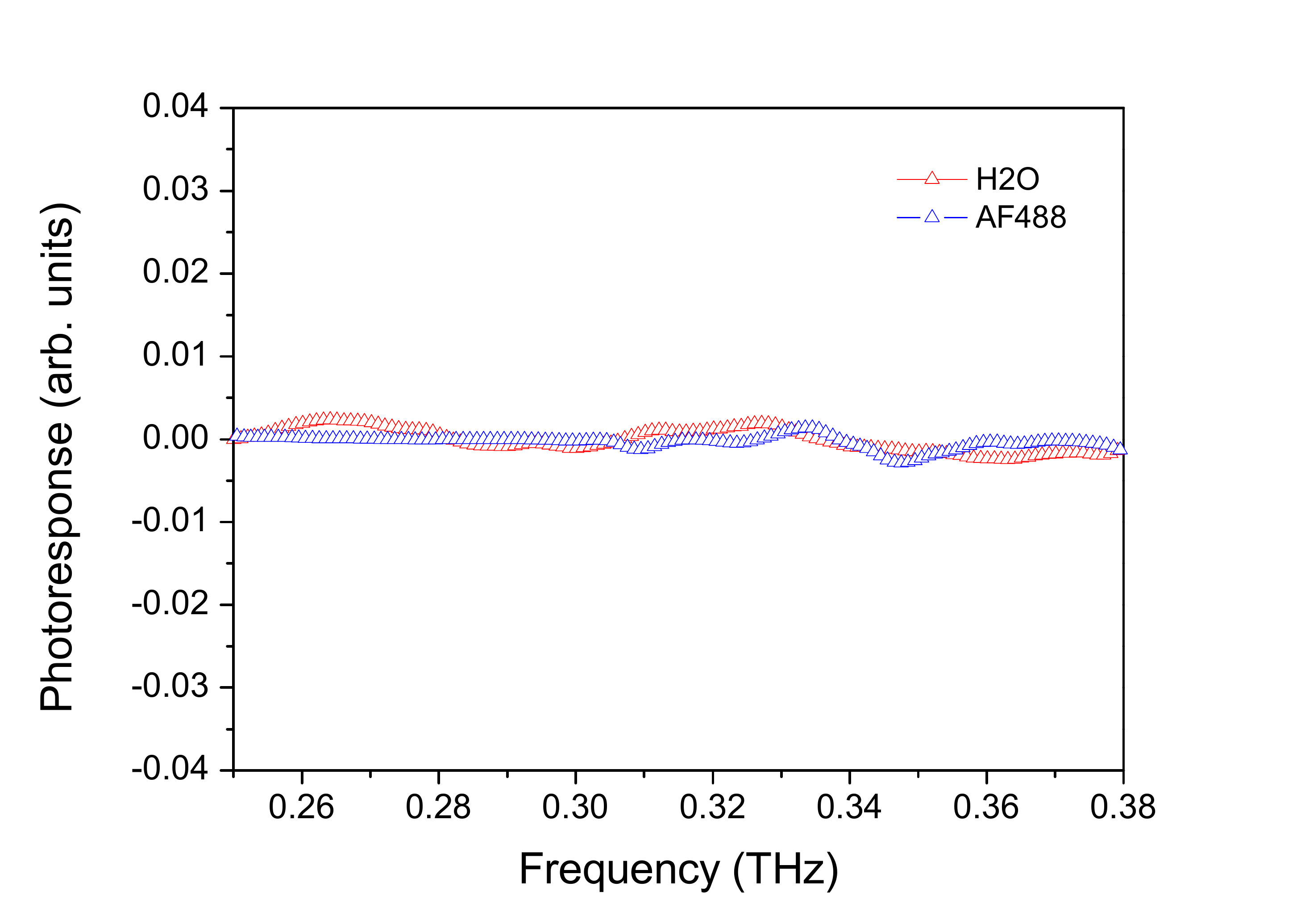} 
\caption{Control data of the rectenna-based absorption spectra. Difference spectra obtained with and without blue laser illumination of water (red triangles), and of a aqueous solution of AF488 fluorochrome at a concentration of 5 mg/mL  (blue triangles). The spectra have been recorded after 10 minutes of blue light illumination.}
\label{AFspectra}
\end{figure}

\subsection{Spectroscopic detection of the collective mode}

\begin{figure}[h!]
\centering
\includegraphics[scale=0.25,keepaspectratio=true]{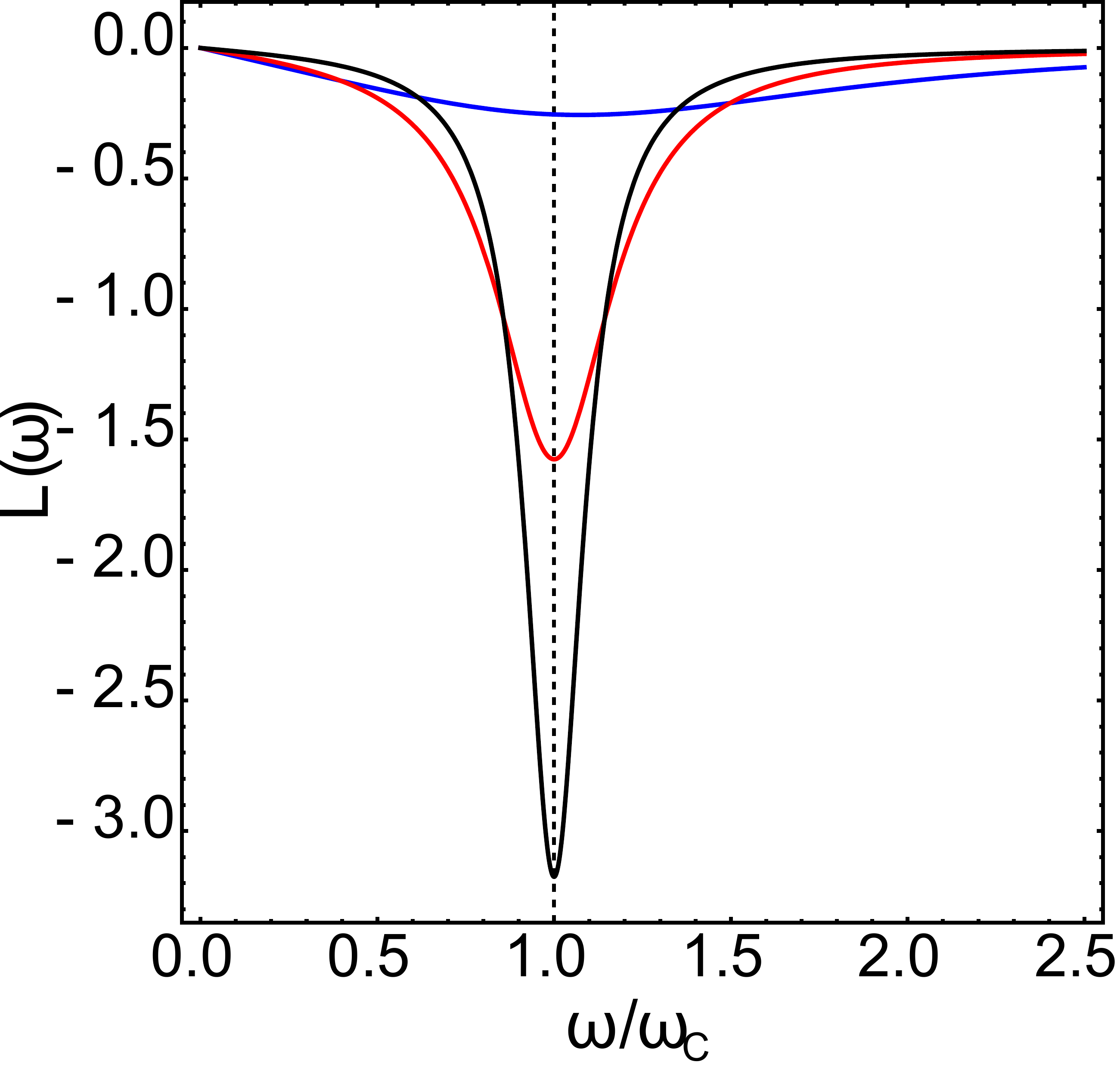} 
\caption{The function $L (\omega)$ of Eq.\protect\eqref{work} is plotted in arbitrary units for three different values of $\tau$. The blue line corresponds to $\tau =1$, the red line to $\tau =5$, the black line to $\tau =10$, having set $\omega =10$. }
\label{Lomega}
\end{figure}

In both experiments the collective oscillation of the BSA protein is seen as a spectroscopic absorption feature. At variance with standard absorption spectroscopy, where the radiation entering the absorbing medium is responsible for the creation of atomic or molecular excited states, in the experiments reported here the THz radiation is used to detect an already excited state of the molecules. Actually, the collective oscillation of the proteins makes them behave as very small antennas (Hertzian dipoles) with the characteristic property of antennas of simultaneously absorbing and emitting electromagnetic radiation. However, the absorption along the THz optical path cannot be compensated by the radiation emitted by the oscillating dipoles because it spreads over all the directions in space. The net result is an absorption feature. 
If we denote with $\vec\mu (t)$ the dipole moment of a protein and with $\vec E(t) = \vec E_0 \cos (\omega t)$ the electric field of the THz radiation, the attenuation of the electric energy density within the drop of protein solution is proportional to the  work done by the electric field, that is, $L = - \vec\mu (t)\cdot \vec E(t)$. The oscillation of the dipole moment is necessarily damped, predominantly because of bremsstrahlung emission, so that, denoting by $\tau$ the lifetime of the activated collective oscillation and by $\omega_{c}$ its frequency, we can set $\vec\mu (t) = \vec\mu_0 e^{-t/\tau}\cos (\omega_{c} t)$. Thus, after averaging over all the relative orientations and all the phase differences $\phi$ between $\vec\mu (t)$ and $\vec E(t)$ such that the electric field does a positive work, we obtain
\begin{equation} 
L (\omega) = 2 \int_0^\pi d\phi \int_0^\infty dt \mu_0 E_0 e^{-t/\tau}\omega_{c} \sin (\omega_{c} t) \cos (\omega t + \phi ) \ .
\label{work}
\end{equation} 
This is the elementary attenuation process of the THz radiation. This process is repeated in time for each molecule at a rate proportional to the intensity of the drop illumination with the blue light. Moreover, the total attenuation is proportional to the concentration of absorbing molecules in the protein solution. 
Equation \eqref{work} gives for $L (\omega)$ a Lorentzian shape centered at $\omega_{c}$, the resonance frequency of the collective oscillation of the BSA protein. Figure \ref{Lomega} shows three different shapes of the function $L(\omega)$  obtained for different values of $\tau$ (in arbitrary units). These line shapes show that an already vibrating underdamped dipole absorbes the weak terahertz-radiation-probe with the same frequency pattern of the $0.314$ THz absorption line reported in Figure 3 of the main text. The latter is well fitted by the $L (\omega)$ function by using a $Q$ factor of $50$ ($Q =\Delta\nu/\nu$, that is the ratio between the line width and the line frequency). 

In Figures \ref{laser-off}a and \ref{laser-off}b some control spectra are reported which have been obtained with the microwire antenna. The two groups of spectra refer to the blue light emitting laser switched off and on, respectively.  The observed absorption line at $0.314$ THz is clearly found only when the fluorochrome AF488 is bound to the BSA molecules and in presence of $488$ nm laser light illumination of the protein solution. These spectra rule out any other origin of the observed absorption feature beside the proposed one in the main text of the present work.

In Figure \ref{romespectrum}  the photoresponse spectra obtained with the rectenna probe highlight the same phenomenology: the absorption line at $0.314$ THz is present only when the aqueous solution of AF488-labelled BSA molecules is illuminated with $488$ nm laser light. A clear difference is again observed between the spectra when the blue light illumination is switched off and on. Artefacts possibly due to blue light illuminated water alone, or possibly due to the blue light illuminated AF488 dye in aqueous solution, are ruled out by the control spectra reported in Figure \ref{AFspectra}, and are in agreement with the analogous controls performed with the microwire antenna reported in Figures \ref{laser-off}a and \ref{laser-off}b.

\subsection{Experimental activation time scale of the collective mode}
In both cases of Montpellier and Rome experiments, the absorption feature, which is attributed to the activation of an out-of-equilibrium collective oscillation of the protein, appeared after several minutes of blue light irradiation: about $4$ minutes in the Montpellier experiment, after $10$ minutes in the Rome experiment (though in this latter case no measurement was performed before $10$ minutes). 
In what follows we provide a consistency estimate of this long activation time with the physical conditions of the experiments. Going beyond a qualitative estimate is hardly feasible, and in any case would not affect the meaning of the experimental outcomes reported in the present work.
Thus we proceed by estimating the balance between the energy input rate and the energy loss rate of each molecule. As we shall discuss at the end of this Section,  this long time  is not an hindrance to the biological relevance of the phenomenon reported in the present work.

An elementary account of the balance between energy gain and loss for each protein can be given by the equation
\begin{equation}
\frac{dE}{dt} = - \frac{2}{3}  \frac{(Z e)^2 |\ddot{\mathbf{x}}|^2}{c^3}  - \Gamma + W
\label{rateequa}
\end{equation}
where $E$ in the l.h.s is the numerical value of the energy of the system described by the Hamiltonian \eqref{eq:HfreeQuan} of the previous Section, the first term in the r.h.s. represents the radiative losses (bremsstrahlung) of the accelerated dipole of moment $Z e \mathbf{x}$, (where $Z$ is the number of elementary charges $e$ and $\mathbf{x}$ is the separation between the center of positive and negative charges) and $W$ is the energy input rate due to the $488$ nm light irradiation. $\Gamma$ stands for other kind of energy losses leading to thermalization of the protein with its aqueous environment.
 
For what concerns the energy input rate $W$, the energy difference between the entering photons, of wavelength $4.88 \times 10^{-5}$ cm, and the fluorescently emitted photons, of wavelength $5.3 \times 10^{-5}$ cm, amounts to $3.18\times 10^{-13} {\rm erg} = 0.19\  {\rm eV}$. When the Argon laser operates at $500$ $\mu$W the number of emitted photons per second is $1.2\times 10^{15}$. The cross section of the fluorochrome Alexa488 is $3.2\times 10^{-16}$ ${\rm cm}^2$ (free dye in water) \cite{lili}, 
so that assuming a Gaussian beam section (at $1/e^2$) of $7.8\times 10^{-3}$ ${\rm cm}^2$ (corresponding to a diameter of about  $0.1$ ${\rm cm}$),  we find that each Alexa488 molecule receives approximately $50$ photons per second. Each protein molecule has an average number of $5 - 6$ fluorochromes bound to it so that, considering that part of them could be partly shadowed by the protein itself, we can assume that the average number of photons received by each protein per second belongs to the interval  $120\div 300$. Hence the upper bound for $W$ is  estimated between $3.8\times 10^{-11}$ erg s$^{-1}$ and $9.5\times 10^{-11}$ erg s$^{-1}$. This estimate is obtained under the assumption that all the  photons hitting the Alexa488 molecules are absorbed; the quantum yield of these dye molecules is high (92 $\%$) so that practically all the harvested photons are converted into fluorescence thus providing an upper bound of the energy available for the protein excitation. Of course part of this energy can be dissipated in the sorrounding environment in the form of heat. 

For what concerns the radiation losses, we proceed to estimate the time average value of 
${\overline{|\ddot{\mathbf{x}}|^2}}$. Since the $63\%$ of the N-isoform of the BSA protein consists of $\alpha$-helices, we can assume that the so-called accordionlike vibrational modes of these helices provide the largest contribution to the protein extension and thus to the variation of the protein's dipole moment. At thermal equilibrium we can assume that the dipole elongation ${\mathbf{x}}(t)$ can be the result of an incoherent superposition like ${\mathbf{x}}(t) = \sum_{i=1}^{30} A_i \cos(\omega_i t + \xi_i)$, where the $\xi_i$ are random phases, and the sum runs over the $30$ $\alpha$-helices of which the protein is composed. To compute the time average of $|\ddot{\mathbf{x}}|^2$ we have to estimate the quantity $\sum_{i=1}^{30}\frac{1}{2} {\overline{A_i^2 \omega_i^4}}$ (since the mixed terms average to zero).
Using the formula $\omega_i = 2\pi (1/2L_i)(E/\rho)^{1/2}$ for the frequencies of the accordionlike modes of the $\alpha$-helices \cite{alphahelix}, where $1/\rho =0.74$, $E = 2.31\times 10^{11}$ dyne cm$^{-2}$ for $\alpha$-helices, $L_i = 1.5 N$ $\mathring{\text{A}}$, where $N$ is the number of amino acid residues, and assuming that $A_i$ belongs to the interval   $0.05 L_i \div 0.1 L_i$, and $Z e = - 13 \times 4.803\times 10^{-10} Fr$ for the BSA protein at neutral pH \cite{barbosa}, we finally find:

\begin{equation}
 \frac{2}{3} \frac{(Ze)^2} {c^3} {\overline{|\ddot{x}|^{2}}}\simeq 3.25\times 10^{-12} \div 1.7\times 10^{-11} {\rm erg\  s}^{-1}\ 
 \label{larmor}
\end{equation}
 
The range of values of the energy input rate and that of the radiative losses are thus almost overlapping. Making these estimates more precise is hardly feasible and is beyond the aims of the present work. What matters here is that since we experimentally observe the activation of the collective mode of the BSA molecules, the energy input rate $W$ must exceed the rate of all the losses, radiative and non, and if this happens only by a small amount, then ${dE}/{dt}$ can be so small as to require some minutes in order to accumulate enough energy in each 
molecule [this can be thought of as a steady increase of the value of the third term of the unperturbed part of the Hamiltonian in Eq.\eqref{eq:HfreeQuan}]. When the energy so accumulated exceeds a critical threshold value, the system undergoes a condensation transition channelling the largest fraction of the input energy into the lowest frequency mode(s). An elementary estimate of the rate of energy losses due to the bremsstrahlung in the condensed phase is obtained by entering into the Larmor formula (\ref{larmor}) the dipole acceleration 
$|\ddot{\mathbf{x}}|$ computed from ${\mathbf{x}}(t) =  A\cos(\omega_C t)$, putting $\omega_C= 0.314$ THz, which is the frequency of the collective mode, and assuming that the dipole elongation $A$ is about $10$ $\mathring{\text{A}}$.  The breakeven between the energy input and the radiative losses is then found for a dipole moment in the range $14500 \div 23000$ Debye, corresponding to an effective number of charges $Z$ approximately in the range $290\div 460$. In other words, in the condensed phase the total dipole of the protein oscillates at a low frequency (with respect to the accordion-like modes of the $\alpha$-helices) so that the collective oscillation can be stable with respect to radiative losses up to the activation of large values of the protein dipole moment.

We have neglected any estimate of the collisional losses  - represented by $\Gamma$ in equation (\ref{rateequa}) - because their estimate as viscous losses - on the basis of Stokes formula - gives an unreasonably high energy dissipation rate with respect to the energy input rate. If this were the case, no collective vibration would be observed at all. There is a vast literature about relaxation phenomena, thus thermalization, of proteins subject to different kinds of radiative excitations. The typical relaxation time scales are in the range of picoseconds to nanoseconds, with a remarkable exception - recently reported in  \cite{lundholm}  - where the thermalization of terahertz photons in a protein crystal happens on a micro- to
milli-second time scale, which is interpreted as being due to  Fr\"ohlich condensation (no matter whether quantum or classical). To the contrary, typical picoseconds to nanoseconds relaxations usually pertain side chains relaxations or small groups of atoms. Also in the case of relaxations of collective modes detected with FIR spectroscopy or with THz Time Domain Spectroscopy  \cite{xie,kneller} the physical conditions are very different with respect to our present experiments. In the mentioned literature the collective modes of proteins are probed at thermal equilibrium, thus in presence of all the other vibrational modes and under the action of weak exciting fields. In our experiments the protein molecules are out-of-thermal equilibrium and are strongly excited through an internal cascade of mode-mode couplings stemming from the fluorescence decay of the attached fluorochromes, and the THz radiation is only used to read the presence of the collective oscillation (see the preceding Subsection).

On the other hand, it has been recently found that the hydration shell of the BSA protein is $25$ $\mathring{\text{A}}$ thick  \cite{susko}, and this seem to be a generic property of solvated proteins  \cite{ebbinghaus}. The microrheology \cite{waigh} of this kind of water-protein system, and in particular its high frequency viscoelasticity, is still an open research subject, making  a quantitative estimate of the term $\Gamma$ in equation (\ref{rateequa}) hardly feasible.

A comment about the prospective biological relevance of the observed phenomenology is in order. The main energy source within living cells is provided by ATP hydrolysis.
The typical intracellular concentration of ATP molecules is given around $1 \ \text{mM}$ implying that a protein molecule in the cell undergoes
around $10^6$ collisions with ATP molecules per second \cite{alberts}. Given the standard free-energy obtained from ATP hydrolysis estimated around $50 \ \text{kJ.mol}^\text{-1} = 8.306\times 10^{-13} \text{erg}$, if we assume that only $1\%$ or $2\%$ of the collisions with ATP will provide energy, a power supply between $8.306\times10^{-9} \text{erg s}^\text{-1}$ and $1.6\times10^{-8} \text{erg s}^\text{-1}$ is potentially available. This could be at least two orders of magnitude larger than the power supplied to each protein in our experiments, but reasonably even much more than two orders of magnitude because we have assumed a hundred percent conversion efficiency of the energy supplied by the laser into internal vibrations of the protein. But this can be hardly the case, thus the condensation mechanism {\it in vivo} can be considerably faster.

\section{Concluding remarks}\label{conclusions}
In the present paper we have studied a classical version of a quantum model put forward many years ago by H. Fr\"ohlich \cite{Frohlich:1968,Froeh:1970,Frohlich:1977}. The classical model displays the same phenomenon of a Bose-like phonon condensation of the normal vibrational modes of a macromolecule that was predicted by the original quantum model. Even though our classical model (just like its quantum predecessor) is very idealized, that is, even if our model cannot be quantitatively predictive, it can nevertheless qualitatively catch the presence of a robust and generic phenomenon undergone by open systems with internal nonlinear interactions: a non-equilibrium phase transition when a control parameter exceeds a critical threshold value. This kind of phase transition is found to correspond to a channelling into the lowest frequency mode of almost all the energy pumped into the system, that is to a collective global molecular vibration. This theoretical result has motivated an experimental - and successful - attempt at confirming it. So in the second part of the present paper we have reported - for a model protein - the unprecedented observation of an out-of-thermal-equilibrium collective oscillation which is in qualitative agreement with the theoretical model and in excellent quantitative agreement with the theoretically expected value of a spheroidal collective vibration of the whole protein. 

The novelty is not the collective oscillation in itself, because several terahertz spectroscopic studies have reported about collective modes of proteins, but all of these studies were performed at thermal equilibrium and mainly carried on using dry or low-hydrated powders because of the very strong absorption of water \cite{turton,acbas,falconer,markelz}, even though more recent studies also addressed solvated proteins \cite{xu,ebbinghaus}. The novelty of both our theoretical and experimental contributions, let us stress this point again, consists of considering \textit{out of thermal equilibrium} conditions. On the other hand, recent studies on solvated BSA in THz \cite{bye} and sub THz frequency range have shown \cite{susko} broad resonances due to an efficient coupling of low frequency modes of the protein with the surrounding water,  and though all of these works were performed at \textit{thermal equilibrium},  in common with these previous studies the experimental part of our present work confirms the relevance of the coupling of the protein with the surrounding water molecules. In fact, the strong absorption feature that we observed in a aqueous solution of the BSA protein put out of thermal equilibrium, reveals that the protein vibrating in its collective mode has to be dressed by ordered layers of water molecules in order to attain an effective dipole moment sufficiently large to overcome  the strong absorption of bulk water.

We anticipate that the theoretical and experimental sides of the work presented in this paper could open a broad domain of systematic investigations about out-of-equilibrium activation mechanisms and properties of collective oscillations of different kinds of biomolecules. Furthermore, as already mentioned in the Introduction, the possibility of exciting out-of-thermal-equilibrium collective oscillations of macromolecules is specifically interesting as a necessary condition to activate resonant long-distance electrodynamic intermolecular interactions  \cite{Preto:2015}. Thus our results explain why electrodynamic interactions between biomolecules have hitherto eluded detection, in fact no attempt has ever been done to detect them by involving biomolecules vibrating out-of-equilibrium. Consequently, our work also motivates new efforts to detect these electrodynamic intermolecular interactions \cite{nardecchia1,nardecchia2}. 

\section*{Acknowledgments}
 We warmly thank Jack A. Tuszynski, Michal Cifra and Anirban Bandyopadhyay for useful discussions ; A. Penarier, T. Cohen and F. Cano for their help in the simulation and the realization of the micro-wire local probes. The project leading to this publication has received funding from the Excellence Initiative of Aix-Marseille University - A*Midex, a French ``Investissements d'Avenir'' programme. This work was also financially supported by the Seventh Framework Programme for Research of the European Commission under FET-Proactive grant  TOPDRIM (FP7-ICT-318121), by the projects SIDERANT and NEBULA financed by the french CNRS, by the Occitanie Region and by Montpellier University through its TOP platform and NUMEV LabeX. 
\medskip

{\bf Author contributions} 
I.N., M.L. and J.T. performed the experiments in Montpellier, J.T. designed and built the experimental setup in Montpellier with the support of P.N., V.G. and M.O. performed the experiments in Rome, the classical formulation of Fr\"ohlich condensation stems from the PhD thesis of M.G., J.P. has also contributed the theoretical part, I.D. did some numerical work, J.S. contributed on the biophysical and biochemical aspects of the work, J.T. and L.V. contributed to conceive the Montpellier experiments and participated to the supervision of all the development of the project. M.P., in quality of project leader supervised and intervened in all the aspects the project.
All the authors contributed to the discussion and to the analysis of the results. M.P. wrote the paper with the help of M.G., M.O., J.S., J.T., L.V., and J.T. made all the figures.


\begin{thebibliography}{99}
\bibitem{classicalcondens} In a classical context, the expression ``Bose-like condensation'' is used to mean that the largest part of the energy fed to a macromolecule is channeled into its lowest frequency normal mode.
\bibitem{nota}This formulation is equivalent to the Schr\"odinger representation in quantum mechanics.
\end{thebibliography}
\end{document}